\documentclass[twocolumn,times]{aastex62}

\usepackage{newtxtext,newtxmath}
\usepackage{graphicx,color}
\usepackage{amsmath,amsfonts,amssymb,cancel,ulem}

\usepackage{hyperref}
\definecolor{NavyBlue}{rgb}{0.0,0,0.5}
\definecolor{Burgundy}{rgb}{0.5,0.0,0.125}
\hypersetup{
    breaklinks=true,
    colorlinks=true,
    citecolor={NavyBlue}, 
    linkcolor={NavyBlue},
    urlcolor={NavyBlue}
}

\def\apj{Astrophys.\ J.}

\def\mnras{Mon.\ Not.\ R.\ Astron.\ Soc.}
\def\aap{Astron.\ Astrophys.}
\def\apjl{Astrophys.\ J.\ Lett.}

\def\physrep{Phys.\ Rep.}
\def\pre{Phys.\ Rev.\ E}
\def\prl{Phys.\ Rev.\ Lett.}

\def\apjs{ApJS}

\def\nat{Nature}

\def\araa{Ann.\ Rev.\ Astron.\ Astrophys.}

\newcommand{\lb}{\ell_b} 
\newcommand{\LS}{\rm LS}

\renewcommand{\Re}{\text{Re}} 
\newcommand{\Rm}{\text{Re}_\mathrm{M}} 
\newcommand{\Rmc}{\text{Re}_\mathrm{M}^{\text{(crit)}}}    

\renewcommand{\vec}[1]{\mathbf{#1}}	
\newcommand{\dd}{\mathrm{d}}        

\newcommand{\dsp}{\displaystyle}

\newcommand{\cm}{\,{\rm cm}}    
\newcommand{\km}{\,{\rm km}}    
\newcommand{\m}{\,{\rm m}}      
\newcommand{\pc}{\,{\rm pc}}     
\newcommand{\kpc}{\,{\rm kpc}}  
\newcommand{\Mpc}{\,{\rm Mpc}}  

\newcommand{\g}{\,{\rm g}}      
\newcommand{\Msun}{\,{\rm M}_{\odot}} 

\newcommand{\GHz}{\,{\rm GHz}}  
\newcommand{\MHz}{\, {\rm MHz}} 
\newcommand{\s}{\,{\rm s}}      
\newcommand{\yr}{\,{\rm yr}}    

\newcommand{\kms}{\km\s^{-1}}    

\newcommand{\mkG}{\,\mu{\rm G}} 
\newcommand{\muG}{\,\mu{\rm G}} 

\newcommand{\erg}{\,{\rm erg}}  
\newcommand{\K}{\,{\rm K}}      

\newcommand{\rad}{\,{\rm rad}} 

\newcommand{\brms}{\,b_{\rm rms}}
\newcommand{\urms}{\,u_{\rm rms}}

\renewcommand{\ne}{n_{\rm e}}

\newcommand{\RM}{\text{RM}}
\newcommand{\DP}{\text{DP}}

\newcommand{\kf}{k_\mathrm{F}}
\renewcommand{\i}{{\mathrm{i}}}

\newcommand\Eq[1]{Eq.\,\ref{#1}}
\newcommand\Fig[1]{Fig.~\ref{#1}}
\newcommand\Sec[1]{Sec.~\ref{#1}}
\newcommand\Tab[1]{Table~\ref{#1}}

\graphicspath{{./}{./}}

\shorttitle{Magnetic fields in elliptical galaxies}
\shortauthors{Seta et al.}

\usepackage{xspace}
\newcommand{\glf}{\textsc{galform}\xspace}
\newcommand\rev[1]{#1}

\begin{document}

\title{\uppercase{Magnetic fields in elliptical galaxies:\\
an observational probe of the fluctuation dynamo action}}

\correspondingauthor{Amit Seta}
\email{amit.seta@anu.edu.au}

\author{Amit Seta} 
\affil{Research School of Astronomy and Astrophysics, 
Australian National University, Canberra, ACT 2611, Australia}
\affiliation{School of Mathematics, Statistics and Physics, 
Newcastle University, Newcastle Upon Tyne, NE1 7RU, UK}

\author[0000-0002-3860-0525]{Luiz Felippe S. Rodrigues}
\affiliation{Department of Astrophysics, Radboud University, Heyendaalseweg 135, 6525 AJ Nijmegen,  Netherlands
}
\affil{School of Mathematics, Statistics and Physics, 
Newcastle University, Newcastle Upon Tyne, NE1 7RU, UK}

\author{Christoph Federrath}
\affiliation{Research School of Astronomy and Astrophysics, 
Australian National University, Canberra, ACT 2611, Australia}

\author[0000-0002-3733-2565]{Christopher A. Hales}
\affil{National Radio Astronomy Observatory, P.O. Box 0, Socorro, NM 87801, USA}
\affil{School of Mathematics, Statistics and Physics,
Newcastle University, Newcastle Upon Tyne, NE1 7RU, UK}


\begin{abstract}
Fluctuation dynamos are thought to play an essential role in magnetized galaxy evolution, saturating within $\sim0.01$~Gyr and thus potentially acting as seeds for large-scale mean-field dynamos. However, unambiguous observational confirmation of the fluctuation dynamo action in a galactic environment is still missing. This is because, in spiral galaxies, it is difficult to differentiate between small-scale magnetic fields generated by a fluctuation dynamo and those due to the tangling of the large-scale field. We propose that observations of magnetic fields in elliptical galaxies would directly probe the fluctuation dynamo action. This is motivated by the fact that in ellipticals, due to \rev{their} lack of significant rotation, the conventional large-scale dynamo is absent and the fluctuation dynamo is responsible for controlling the magnetic field strength and structure. By considering turbulence injected by Type Ia supernova explosions and possible magnetic field amplification by cooling flows, we estimate expected magnetic field strengths of $0.2~\text{--}~1 \,\mu{\rm G}$ in the centers of quiescent elliptical galaxies. We use a semi-analytic model of galaxy formation to estimate the distribution and redshift evolution of field strengths, tentatively finding a decrease in magnetic field strength with decreasing redshift. We analyse a historical sample of radio sources that exhibit the Laing-Garrington effect (radio polarization asymmetry in jets) and infer magnetic field strengths between $0.14~\text{--}~1.33 \,\mu{\rm G}$ \rev{for a uniform thermal electron density and between $1.36~\text{--}~6.21\,\mu{\rm G}$ for the thermal electron density following the King profile}. We examine observational techniques for measuring the magnetic field saturation state in elliptical galaxies, focusing on Faraday rotation measure grids, the Laing-Garrington effect, synchrotron emission, and gravitational lensing, finding appealing prospects for future empirical analysis.
\end{abstract}




\section{Introduction} \label{intro}

Magnetic fields are ubiquitous in galaxies and their presence is usually explained by the turbulent dynamo theory, a process by which the turbulent kinetic energy in the medium is converted into the magnetic energy 
\citep{BS2005,Fed16,Rincon19}. Fluctuation dynamo, a type of magnetohydrodynamic (MHD) dynamo, is produced by a turbulent flow (characterized by the hydrodynamic Reynolds number $\Re = \urms \ell_0/\nu$,
where $\urms$ is the root-mean square velocity field, $\ell_0$ is the driving scale of the turbulence and $\nu$ is the viscosity) and therefore fluctuates in time and space. 
A weak seed magnetic field in a turbulent electrically conducting fluid either grows exponentially or decays. The field grows only when the magnetic Reynolds' number ($\Rm = \urms \ell_0/\eta$, where $\eta$ is the magnetic resistivity)
is greater than its critical value, $\Rmc$. For fluctuation dynamos, the $\Rmc$ is approximately around $100$ \citep[can be higher for highly compressible and supersonic flows;][]{Fed14} whereas the estimated $\Rm$ in spiral galaxies is of the order of $10^{18}$ \citep{BS2005}.
Thus, the fluctuation dynamo is expected to be active in galaxies.
Initially, the field is weak and the effect of the magnetic field on the velocity flow is negligible 
(this is referred to as the kinematic fluctuation dynamo). When the field grows, the magnetic field becomes strong enough to react back on the flow via the Lorentz force. This alters the properties of the turbulent flow. 
Eventually, the magnetic field saturates due to the back reaction of the Lorentz force on the flow. 

MHD theories predict that a fluctuation dynamo will arise practically in any turbulent medium (as long as $\Rm > \Rmc$). The negligible seed field in protogalaxies is amplified quickly (in about $10^7 \yr$ in spiral galaxies) to magnetic field strengths (of the order of $1\text{--}10\mkG$)  in near equipartition with the turbulent kinetic energy. The fluctuation dynamo amplified field exists at small scales (with largest scale of the order of 
the driving scale of turbulence, $\ell_0 \approx 100\pc$ in disks of spiral galaxies) and it requires the mean-field dynamo \citep{RSS88,Beck1996,BS2005} to further amplify and  order the field over the observed $\kpc$ scales in spiral galaxies \citep[e.g. in M51;][]{Fletcher2011}. Besides turbulence, the mean field dynamo also requires large-scale galaxy properties such as rotation, velocity shear and density stratification. Fluctuation dynamo plays an important role in seeding the mean field dynamo. \rev{The fluctuation dynamo action is described in detail in \citet{BS2005}.}

Besides fluctuation dynamo, small-scale magnetic fields in spiral galaxies can also be generated by tangling of the mean (or large-scale) field and compression by shocks \citep[for further discussion of these mechanisms, see Appendix A in][]{SSWBS18}. 
The magnetic field generated by compression due to shocks is possibly correlated with the gas density \citep{Fed11,Fed14}.  The small-scale magnetic field due to a fluctuation dynamo is spatially intermittent, whereas that due to the tangling of the mean field is presumably Gaussian in nature due to pervasive Gaussian turbulence. 
Thus, the fluctuation dynamo generated magnetic field also decides the small-scale structure of magnetic fields in galaxies. In this paper, we propose how studying magnetic fields in quiescent elliptical galaxies can serve as a probe of the fluctuation dynamo action and discuss ways about pursuing the same.

Considering the driving scale of the turbulence in the interstellar medium (ISM) of spiral galaxies within a range of $50 \text{--} 100 \pc$ \citep{OS93,Gaenslar2005,Fletcher2011,Houde2013}\footnote{Some observational
studies report even smaller values of $\ell_0$ in the range $1 \text{--} 20 \pc$ \citep{MS96,Haverkorn08,Iacobelli2013}.}, it would require a spatial resolution of $\lesssim10\pc$  to study the structure of small-scale magnetic fields in spiral galaxies.
Such a resolution is extremely difficult to achieve with present-day telescopes and thus it is difficult to observationally differentiate between the magnetic field 
generated by a fluctuation dynamo and that due to the tangling of the mean field.
In a magnetized turbulent environment like the ISM of spiral galaxies, the fluctuation dynamo action is physically the most plausible scenario for magnetic field amplification but 
we are yet to confirm this in a galactic environment. 
The existence of a fluctuation dynamo in galaxies is fundamental to the galactic dynamo theory. In the absence of a fluctuation dynamo, the magnetic field
in protogalaxies would be seeded with a much weaker field and then the mean-field dynamo would take much longer to amplify the field up to the value observed today \citep{RSS88,Arshakian09}. 
Furthermore, the field generated by a fluctuation dynamo is important for low-energy cosmic ray propagation \citep{SSSBW17} and for interpretations
of radio observations. Thus, it is important to find a clear observational probe of the fluctuation dynamo action.

With a motivation to probe the fluctuation dynamo action in a galactic environment, we consider magnetic fields in elliptical 
galaxies (slow rotators, as defined by the ATLAS$^{\rm 3D}$ Project in \citealt{Atlas3d2,Atlas3d3} \footnote{http://www-astro.physics.ox.ac.uk/atlas3d/}) . 
Unlike spirals, elliptical galaxies have a very small rotation rate (with rotation velocity  of about $20 \kms$ as compared to $200 \kms$ in a typical spiral galaxy)
and so a mean-field dynamo (conventional $\alpha \omega$ dynamo) action is unlikely. 
Moreover, these systems are not differentially rotating, so a magnetorotational instability generated large-scale field is also unlikely. 
However, a weak seed magnetic field can still be amplified by
the fluctuation dynamo action due to the interstellar turbulence driven by 
supernova explosions and the random motions of stars and gas 
induced by galactic dynamics \citep{MShu96}.
The resulting magnetic field can be further amplified by compression produced by cooling flows \citep{Fabian94} in elliptical galaxies \citep{MB97}.
This leads to a magnetic field which is non-Gaussian and spatially intermittent.

Thus, the detection and characterization of magnetic fields in elliptical galaxies would be a direct observational confirmation of the fluctuation dynamo action.
Moreover, the observations can be compared with the theories of fluctuation dynamos to further understand the physics of magnetic field saturation,
which is an unsolved problem \citep[a few possible saturation mechanisms are discussed in][]{TCB2011,SBS2015,Rincon19,Seta19}. 
It is also unclear whether the dominant scale (at which the
magnetic power is at its maximum) lies close to the resistive scale
\citep{SCHMM02,SCTMM04} or whether it is in fact much larger \citep{Sub99}.  
The spatially intermittent structure of magnetic fields generated by a
fluctuation dynamo has been studied both analytically \citep{ZRS90} and numerically \citep{Wilkin2007,Seta19} but lacks 
observational confirmation. 

The structure of the paper is as follows. In \Sec{ellip} we discuss turbulence in elliptical galaxies and predict the magnetic field strength generated by that turbulence. In \Sec{cosmo} we predict properties of magnetic fields in elliptical galaxies obtained from a semi-analytical cosmological simulation. In \Sec{RMproperties} we predict properties of the rotation measure distribution and present results from numerical simulations to verify these predictions. In \Sec{ExiObs} \rev{we estimate the magnetic field strengths using the Laing-Garrington effect and also} examine observational prospects for measuring the magnetic field properties in elliptical galaxies using existing and forthcoming radio telescopes. Finally,  we conclude in \Sec{conc}.


\section{Magnetic fields in elliptical galaxies} \label{ellip}

\subsection{Turbulence in the hot gas of elliptical galaxies} \label{ellip1}
The source of hot gas in ellipticals can be internal or external to the galaxy. 
Internally, the hot gas can be due to outflows from evolved stars and Type Ia supernova explosions. Externally, some of the circumgalactic gas 
(expelled from the galaxy by winds driven by frequent Type II supernovae explosions at earlier stages of galactic evolution) 
can fall back into the galaxy, as well as the gas accreted from the galaxy group or cluster.
The hot gas in elliptical galaxies is observed via its X-ray emission. The relationship between the X-ray luminosity $L_X$ and optical luminosity (in B band) $L_B$ of bright elliptical
galaxies is observationally found to be $L_X \propto L_B^{2.2}$  \citep{OSullivan2001}. This confirms the non-stellar origin of the X-ray emission, otherwise the relationship would have been linear.
From X-ray observations, gas mass can be inferred. The total mass of the hot gas is about $1\text{--}2\%$ of the total stellar mass \citep{MB03}. 

The turbulence in spiral galaxies which merge to form an elliptical galaxy decays. The numerical simulations of decaying three-dimensional isotropic turbulence in an isothermal gas
suggests that the turbulent velocity $\urms$ decays as a power-law, $(t/t_0)^{-\eta_u}$, with the exponent ${\eta_u}$ in the range $0.4\text{--}0.6$ and $t_0$ being the eddy turnover time \citep{MacLow98,Stone98,SSH06,Sur19}.
Considering $\eta_u$ as $0.6$, the rms turbulent velocity in the parent spiral galaxy, $\urms\simeq10\kms$ decreases to $1 \kms$ in $5\times10^{8} \yr$ 
(where the eddy turnover time is calculated as follows: $t_0 \approx \ell_0/\urms \approx 100\pc/10\kms \simeq10^{7} \yr$). However, we aim to look for magnetic fields in quiescent elliptical galaxies, which
probably had a major merger a few Gyr earlier, so the turbulence due to parent spirals must have significantly decayed by then. Also, there is a lower level of turbulence (as compared to the parent spiral galaxies) 
maintained in quiescent elliptical galaxies due to continuous energy injection by Type Ia supernovae explosions. This is also seen in observations. Each explosion of a Type Ia supernova enriches the medium with 
roughly $0.7 \Msun$ of iron and iron abundance in the hot gas of ellipticals increases from the outskirts to the centre of the galaxy, where the density is highest, so the supernova activity
is highest \citep{MB03}. Further, the observed iron abundance can be used to estimate the supernova explosion rate in ellipticals ($10^{-3} \yr^{-1}$).
The rates are a factor of $\sim3$ lower than in spiral galaxies ($ 3 \times10^{-2} \yr^{-1}$). Thus, the turbulence decays immediately after the merger of spiral galaxies but then is maintained at a certain level
by continuous energy injection by Type Ia supernova explosions in quiescent elliptical galaxies. 
Type Ia supernova explosions will drive irrotational turbulence (negligible vorticity) in the ISM of elliptical galaxies. But this irrotational turbulence
interacts with the multiphase ISM (see ~\citealt{Miao2019}, and references therein, 
for reasons justifying the presence of inhomogeneous, multiphase gas in ellipticals) to generates vortical motions required for the fluctuation dynamo action.

To estimate the properties of turbulence in elliptical galaxies, we closely follow \cite{MShu96}\footnote{The
method is also very similar to that in \cite{Shukurov2004} and \cite{MLK04} where turbulent properties of the interstellar medium
are estimated for supernova explosions in spiral galaxies.} 
with two major changes. First, they consider two drivers of turbulence: supernova explosions and random star motions,
whereas we 
only consider Type Ia supernova explosions since the length scales associated with random star motions would be much smaller than those due to supernova explosions. 
Secondly, they consider acoustic turbulence but we consider hydrodynamic turbulence, because \cite{MB97} showed that the acoustic modes will quickly decay and cannot maintain the pervasive turbulence.

\subsection{Estimating the properties of the ISM turbulence in elliptical galaxies}  \label{ellip2}
X-ray observations suggest a temperature $T$ of the order of $10^7 \K$ for the diffuse interstellar gas in elliptical galaxies \citep{MB03}. 
This implies a sound speed $c_s \approx \sqrt{k_BT/m_\text{p}}\approx300 \kms$, where $k_B=1.38\times10^{-16} \erg \K^{-1}$ is 
the Boltzmann constant and $m_\text{p}=1.67\times10^{-24} \g$ is the proton mass (assuming that most of the hot gas mass is dominated by protons) . 
We now use the sound speed and Sedov--Taylor blast wave self-similarity solution \citep{OM88} to
calculate the driving scale of the turbulence. The radius of a supernova remnant $R$ as a function of time $t$ is then
\begin{equation}
R = \left(\kappa \frac{E_{\rm SN}}{\rho_0}\right)^{1/5} t^{2/5},
\end{equation}
where $E_{\rm SN}$ is the energy of a supernova explosion, $\rho_0$ is the average gas density of the ambient medium and $\kappa \sim 2$ for a monoatomic ideal gas.
Now, the radius $R$ is roughly equal to the driving scale of the turbulence $\ell_0$, when the velocity of the shock front is equal to $c_s$. Thus, $t \approx R/c_s$. This gives
\begin{equation} \label{l01}
\ell_0 \simeq \left(\kappa \frac{E_{\rm SN}}{\rho_0}\right)^{1/3} \frac{1}{c_s^{2/3}}.
\end{equation}
Using $E_{\rm SN}=10^{51} \erg$, 
$\rho_0=m_\text{p}\,n_0 ={1.67\times10^{-27}\g \cm^{-3}}$ (assuming an ambient number density, $n_0=10^{-3} \cm^{-3}$, \citealt{Forman85}) 
 and $c_s=300 \kms$, we obtain $\ell_0\simeq300 \pc$. 
 This scale is slightly larger than that in spiral galaxies ($\simeq 100 \pc$) because the ambient medium is less dense.

The supernovae survive longer in ellipticals as compared to the spirals, but their frequency is much lower and so they inject less turbulent kinetic energy in the medium. To calculate the turbulent velocity,
we assume that about $1\%$ of the total supernovae energy is converted to the kinetic energy of the turbulence \citep{DW97}. The rate at which supernovae inject energy into the medium can be balanced
with the rate of gain of kinetic energy of the medium as
\begin{equation} \label{v01}
\frac{v_0^2}{\ell_0/v_0} \approx f  \nu_{\rm SN} E_{\rm SN} M_{\rm gas}^{-1}, 
\end{equation}
where $v_0$ is the turbulent velocity, $f$ is the fraction of energy that is injected into the medium, $\nu_{\rm SN}$ is the frequency of supernova explosions
and $M_{\rm gas}$ is the mass of the hot gas. Using $\ell_0=300\pc$, $f=0.01$, 
$\nu_{\rm SN}=10^{-3} \yr^{-1} $\citep{capp99}, $E_{\rm SN}=10^{51} \erg$ and $M_{\rm gas}=10^{10} \Msun$, we obtain $v_0 \simeq 2.5 \kms$.
This is lower than that in spiral galaxies ($\simeq 10 \kms$). 

The estimated driving scale of turbulence ($300 \pc$) is roughly similar to that obtained in \cite{MShu96} ($350 \pc$), but the turbulent velocity we get ($2.5 \kms$) is smaller than their estimate ($60 \kms$).
This is because we consider that Type Ia supernova explosions drive hydrodynamic turbulence rather than the acoustic turbulence, which they assume. The acoustic turbulence decays very quickly due to 
relatively high viscosity in the ISM of elliptical galaxies. For a fully ionised gas, with electron number density $\ne$ and temperature $T$, the viscosity $\nu$ based on the kinetic theory is estimated as \citep{BS2005}
\begin{align}
\nu \simeq 6.5 \times 10^{22}  \left(\frac{T}{10^{6} \K}\right)^{5/2} \left(\frac{\ne}{1 \cm^{-3}}\right) \cm^2 \s^{-1},
\label{nu}
\end{align}
where we assumed Coulomb logarithmic factor of order unity. For the core for an elliptical galaxy, $\ne = 0.1 \cm^{-3}$ and $T = 10^{7} \K$, we obtain $\nu \simeq 2 \times 10^{26} \cm^2 \s^{-1}$. 
The diffusive length scale $l_{\nu} = (\nu/t)^{1/2}$ in an eddy turnover time $t_0 = 300 \pc/2.5 \kms \simeq 10^{7} \yr$ is $3 \times 10^{5} \cm$, which is much smaller than the driving scale ($\sim 300 \kpc$).
Thus, high viscosity dissipates acoustic waves within a very short distance. Plasma effects due to the presence of magnetic fields can reduce the viscosity calculated via \Eq{nu}, but not significantly
to account for such a large difference. Furthermore, the dissipation of acoustic waves might provide an additional heating mechanism for the medium \citep[details of mechanism in][]{Fabian2005, Zweibel2020}.

The driving scale of turbulence $\ell_0\simeq300\pc$ and the rms velocity $v_0\simeq2.5\kms$ are global values calculated from average quantities close to the core of a typical elliptical galaxy\footnote{The high values of turbulent velocities of around $100\kms$ reported using X-ray observations \citep{PZ2012,OZ2017} are for a much larger halo of the elliptical galaxy. Such high values of turbulent velocities can be due to interaction with the AGN at the center of an elliptical galaxy or the surrounding environment.}. 
 The eddy turnover time can be calculated as $t_0 \approx \ell_0/v_0\simeq1.2\times10^{8} \yr$.
The typical value of magnetic Reynolds numbers can also be calculated.
Spitzer resistivity $\eta$ for a plasma at a temperature $T$ can be estimated from 
\citep{BS2005}:
\begin{align}
\eta & \simeq 10^{4} \left(\frac{T}{10^{6} \K}\right)^{-3/2} \cm^2 \s^{-1},
\end{align}
where we assumed the Coulomb logarithmic factor of order unity.
Using $T=10^{7} \K$, we obtain $\eta  \simeq 3 \times 10^{2} \cm^2 \s^{-1}$.
For $\ell_0=300\pc$ and $v_0=2.5\kms$, we obtain $\Rm = \ell_0v_0/\eta \simeq 10^{24}$. The estimated $\Rm$
is considerably greater than $\Rmc$ ($\approx10^{2}$), which is required for the fluctuation dynamo action.
Even though the flow is very viscous, such a high magnetic Reynolds number ensures fluctuation dynamo action in random flows. Thus, it is reasonable to expect fluctuation dynamo-generated magnetic fields in elliptical galaxies.

Since the gas and stellar densities in elliptical galaxies vary with radius (usually the variation is described
by the King profile with a central core radius, see \citealt{St12})
the properties of the ISM turbulence would also change. 
We, however, are interested mostly in the core region of the elliptical galaxies where the density of the hot gas and supernova activity are higher 
and thus the expected magnetic field strength and the probability for an unambiguous detection is also higher.
The magnetic field strength may be further enhanced close to the core due to the compression by cooling flows.

\subsection{Fluctuation dynamo action and further amplification by cooling flows} \label{ellip3}
The rms value of fluctuation dynamo generated magnetic field $\brms$ is in near equipartition with the turbulent kinetic energy \citep{Haugen2004,Fed11,Tz18,Seta19},
\begin{equation}
\brms \approx 0.5 (4 \pi \rho)^{1/2}v_0,\label{eq:beq}
\end{equation}
where $\rho$ is the density of the medium and $v_0$ is the turbulent velocity. 
This is probably representative of magnetic field strengths in the core of elliptical galaxies.
In the core of a typical elliptical, with $\rho=0.1m_\text{p}=1.67\times10^{-26}\g \cm^{-3}$ \citep{Forman85} and $v_0=2.5\kms$ (as obtained in \Sec{ellip1}), we obtain $\brms\simeq0.2\mkG$.

The gas in the centers of elliptical galaxies quickly loses its energy via X-ray emission. Since the central core cools quickly, the weight of the outer region will cause the surrounding material to flow inwards. This is referred to as the cooling flow \citep{Fabian94}. They
can also be inferred from the observations of O $\rm VI$ line emission from elliptical galaxies \citep{Bregman01,Bregman05}.
The magnetic field would be further enhanced by such cooling flows due to compression. By flux conservation under spherically symmetric compression, the field strength $b$ grows as $\rho^{2/3}$, where $\rho$ is the gas density. In cooling flows, the density can be enhanced by a factor of $10$ or more \citep{MB97}. Thus, the magnetic field with magnitude approximately equal to $0.2\mkG$ can be further amplified to be around $1\mkG$. \cite{MB97}, using a detailed spherical cooling flow model, suggest a radial dependence for the magnetic field strength of the form $b(r) \propto (1\text{--}10) \, r^{-1.2} \mkG$. The radial dependence of thermal electron number density is $\ne(r) \propto 0.1 r^{-1.5} \cm^{-3}$ \citep{MB03}. Thus, we do expect a significant Faraday rotation of the polarized light from the background radio sources passing through the elliptical galaxy in the front (further discussed in \Sec{RMproperties} and \Sec{RMG}).


\section{Predictions from Cosmological Simulations}\label{cosmo}

We used the \glf semi-analytic model of galaxy formation \citep{Lacey2016} to estimate the expected distribution of magnetic field strengths in elliptical galaxies 
and its redshift evolution.  We classify any galaxy as elliptical in \glf's output
where the bulge-to-total luminosity ratio in the $r$-band $(B/T)_r>0.5$ (see \S4.2.3 of \citealt{Lacey2016} and \citealt{Benson2010} for details). For the purposes of the calculations in this section, the contents  of the remainder disk components of the selected galaxies were ignored.

As earlier, we assume that the magnetic field is amplified by a fluctuation dynamo \rev{(no further amplification due to the cooling flows is considered in the present model)}, with a steady-state rms strength given by Eq.~\eqref{eq:beq}. For the sake of definiteness 
we compute the equipartition strength at the half-mass radius, $r_{1/2}$, of each individual elliptical galaxy. Thus, 
\begin{align} 
    \left.b_\text{rms}\right|_{r=r_{1/2}}  \\
            &= 11.6 \muG \left(\frac{M_\text{g}}{10^{10}\Msun}\right)^{1/2}\\
             &\times                 \left(\frac{\sigma_{1/2}}{50\kms}\right)
                             \left(\frac{r_{1/2}}{5\kpc}\right)^{-3/2}\nonumber
\end{align}
where we have assumed that the density follows a \citet{Hernquist} profile, with $M_g$ corresponding to the total gas mass of the elliptical galaxy, and $\sigma_{1/2}$ is the velocity dispersion at the half-mass radius.

\begin{figure}
    \centering
    \includegraphics[width=\columnwidth]{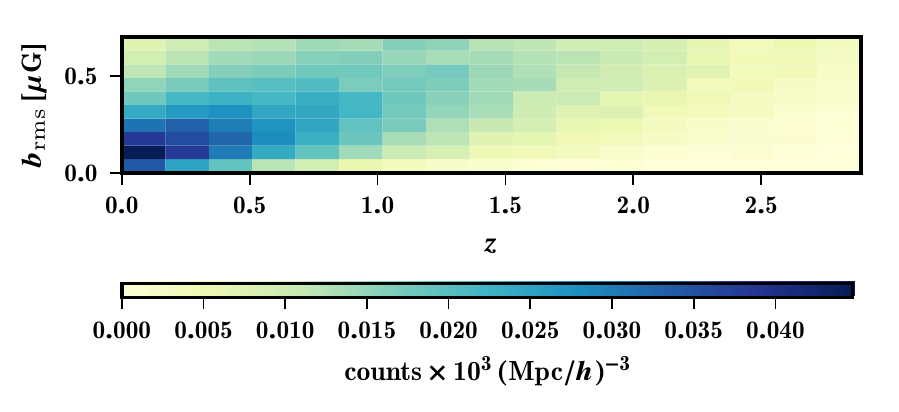}
    \caption{Redshift evolution of the distribution of $b_\text{rms}$ at the
             half-mass radius of simulated elliptical galaxies with 
             non-negligible gas content.
             Colors show the number of galaxies in each $b_\text{rms}$ per comoving volume (where $h$ is the 
             is the Hubble parameter).
             }
    \label{fig:simulated_brms}
\end{figure}

\Fig{fig:simulated_brms} shows the predicted 
distribution of rms strength of the magnetic field at the half-mass radius of elliptical galaxies as a function of redshift (ignoring galaxies without any gas). For the same comoving volume, we find that at a larger redshift the typical rms strength of the magnetic fields in elliptical galaxies is larger than at lower redshifts, whereas the number of counts is expected to decrease.
These trends follow the hierarchical assembly history of galaxies: the total number of elliptical galaxies increases with time, but the amount of available gas (and therefore, the typical field strength) decreases.


\section{Properties of Faraday Rotation Measure Distribution from fluctuation dynamo simulations}\label{RMproperties}

\subsection{Theoretical expectations for standard deviation of rotation measure fluctuations} \label{theory}
\rev{The polarization angle of the linearly polarized synchrotron radiation rotates when the radiation passes through a magnetized medium containing thermal electrons. The amount of rotation is a function of the wavelength of the radiation and the Faraday rotation measure, $\RM$, which} in terms of convenient units is written as 
\begin{equation}
\frac{\RM}{1 \rad \m^{-2}} = K \int_{0}^{L/1 \pc} \frac{\ne}{1 \cm^{-3}} \frac{b_{\parallel}}{1 \mkG} \dd l,
\label{defRM}
\end{equation}
where $\ne$ is the thermal electron number density, $b_{\parallel}$ is the magnetic field parallel to the line of sight, $L$ is the path length, and $K=0.81$ when the units are scaled as indicated. For a cosmologically distant source at a redshift $z$, the $\RM$ in the observer frame is further reduced by a factor of $(1+z)^2$ than that given by \Eq{defRM} (which is the $\RM$ in the rest frame).

The standard deviation of fluctuations in rotation measure $\sigma_\RM$ can be estimated from the two-point correlation function of $\RM$, which in turn can be estimated from the two-point correlation of the magnetic field. For the simplest case, this can be done by assuming that the number density of thermal electrons
is uniform. Assuming a constant $\ne$ in \Eq{defRM}, \cite{SS2008} demonstrated that
\begin{equation}
\frac{\sigma_{\RM}}{1+z^2} = \frac{(2 \pi)^{1/4}}{3^{1/2}} K n_e \brms (L \lb)^{1/2},
\label{sigRM}
\end{equation}
where $\brms$ is the root mean square magnetic field strength in the region, $L$ is the path length and $\lb$ is the correlation length of the random magnetic field. In the central region, \rev{assuming that the fluctuation dynamo is active (see \Sec{ellip3})}, $\ne = 0.1 \cm^{-3}$ \citep{MB03}, $\brms = 1\mkG$, $\lb = 100 \pc$ ($\ell_0/\lb\simeq3\text{--}4$ and $\ell_0\simeq300\pc$), and $L=10 \kpc$ (since the emission is considered close to the central region). For $z\approx0$, these values yield $\sigma_{\RM} \simeq 75 \rad \m^{-2}$. We expect that this can be observed in case of good signal to noise ratio. However, the number density decreases with the radius and this might further decrease the $\sigma_{\RM}$ at larger radii.

The analysis can be extended to include the contribution from a varying thermal electron number density (see \cite{Felten96} and Appendix A in \cite{BS13}). Assuming the King profile for electrons in an elliptical galaxy $\ne = \ne(0)\left(1 + (r/a)^2\right)^{-3/4}$, where $\ne(0)$ is the electron number density at $r=0$, $a$ is the core radius, and also assuming that $\brms \propto \ne^{\gamma}$, $\sigma_{\RM}$ at $r=0$ is
\begin{equation}
	\frac{\sigma_{\RM}(0)}{1+z^2} = \frac{(2 \pi)^{1/4}}{3^{1/2}} K \ne(0) \brms  (a \lb)^{1/2} \left(\frac{\Gamma\left(\frac{3}{2}(\gamma + 1) - 0.5\right)}{\Gamma\left(\frac{3}{2}(\gamma + 1)\right)}\right)^{1/2}.
\label{sigRMne}
\end{equation}
Using $\ne(0)=0.1 \cm^{-3}, \brms=1\mkG, a=3 \kpc, \lb=100 \pc,$ and $\gamma=2/3$  (motivated by the flux freezing or cooling flow constraint, i.e., $b \propto \rho^{2/3}$), we obtain $\sigma_{\RM}(0)\simeq35\rad \m^{-2}$ at $z\approx0$. The $\sigma_{\RM}$ as a function of the radial distance in the plane of the sky $r_{\perp}$ is
\begin{equation}
\sigma_{\RM}(r_{\perp}) = \frac{\sigma_{\RM}(0)}{1+z^2} \left[1 + (r_{\perp}/a)^2\right]^{-\dsp \left(\frac{1}{4}\Gamma\left(\frac{3}{2}(\gamma + 1) - 1\right)\right)}.
\label{sigRMner}
\end{equation}
For $\sigma_{\RM}(0)/(1+z^2)=35\rad \m^{-2}$ and $\gamma=2/3$, we obtain $\sigma_{\RM}(r_{\perp}) = 35\rad \m^{-2}  (1 + (r_{\perp}/a)^2)^{-0.22}$.

Therefore, we expect significant fluctuations in Faraday rotation from at least the core of an elliptical galaxy (usually defined as the radius at which the slope of the surface brightness profile changes dramatically from low values).

\subsection{Numerical simulations of fluctuation dynamo}  \label{ellipmock2}

\begin{figure*}
\centering
\includegraphics[width=2\columnwidth]{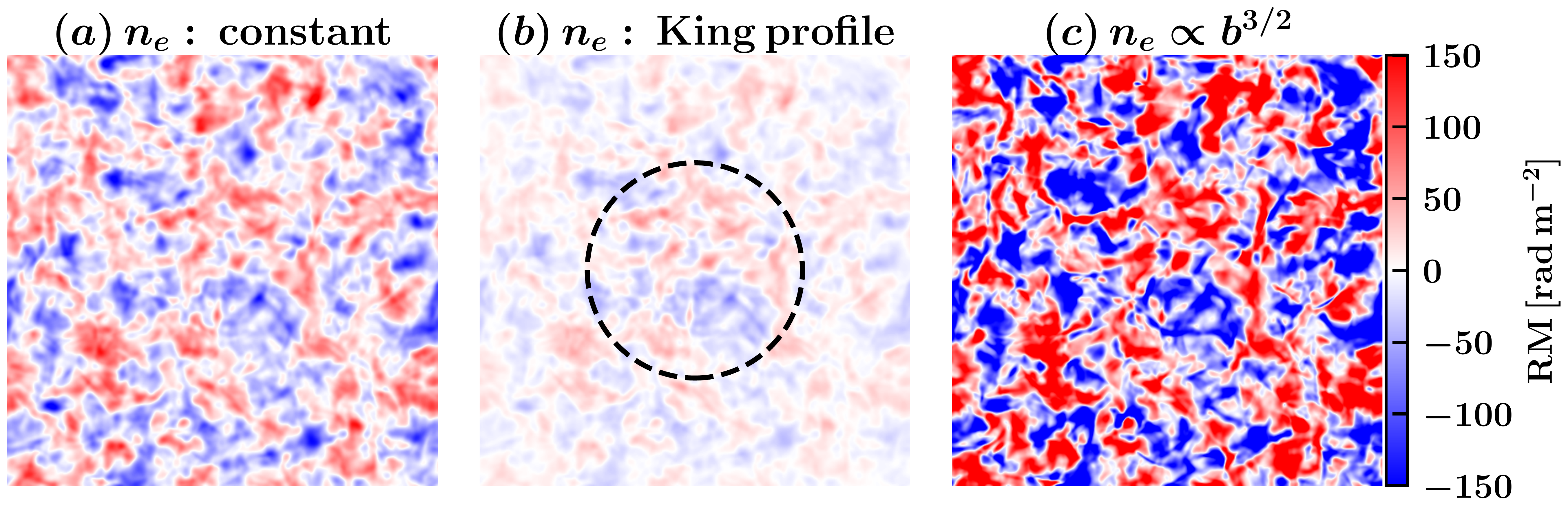}
\caption{Two dimensional $\RM$ maps obtained using the intermittent magnetic field from a fluctuation dynamo simulation and three different electron number density distributions: 
(a) constant $\ne$, (b) $\ne$ following the King profile, $\ne(r) = \ne(0)/(1 + (r/a)^2)^{-3/4}$, where $\ne(0)$ is the number density at the center, $a$ is the core radius ($L/4$), and $r=((x-L/2)^2 + (y-L/2)^2 +(z-L/2)^2)^{1/2}$ is the distance from the center, and (c) $\ne$ from the  flux-freezing or cooling-flow condition, $\ne \propto b^{3/2}$, where $b$ is the field strength. The dotted, black circle in the case (b) shows the core radius $a=L/4$ in the King profile. For all three cases, over the entire numerical domain, the magnetic field is normalized to have a root mean square value of unity ($b/\brms=1$), and the mean thermal electron density is also normalized to unity ($\langle \ne \rangle = 1$). The structures in the cases (a) and (b) are very similar, except that the intensity of $\RM$ decreases as a function of $r$ \rev{in the latter case}. \rev{In case (c), the structures are different at smaller scales.} The dynamo-generated magnetic field is exactly the same in all three cases, and the difference in the \rev{amplitude and the small-scale} $\RM$ structures is due to different thermal electron distributions.}
\label{rm2d}
\end{figure*}
We confirm our expectations for $\sigma_{\RM}$ using numerical simulations of the nonlinear fluctuation dynamo (numerically solving Navier-Stokes with a prescribed random forcing and induction equation for an isothermal gas in a periodic domain as described in \cite{Haugen2004}). The complete details of the simulations are described in Section II of \cite{Seta19}. The forcing is solenoidal, nearly incompressible and the Mach number of the flow is less than $0.1$ We note that in numerical simulations of (Type Ia) supernova-driven turbulence, the overall Mach number is typically $<0.1$ \citep{Miao20192}. The initial conditions of our simulations are as follows: a uniform density, zero velocity, and a very weak random magnetic field with no net flux. As the turbulence is driven numerically within the domain, the density remains roughly the same (since the gas is isothermal, the turbulent driving is near incompressible, and the Mach number is less then unity), and the velocity reaches a statistically steady state after about 2 turbulent eddy turnover times \citep[e.g.,][]{FederrathKlessenSchmidt2009,PriceFederrathBrunt2011}. The magnetic field first grows exponentially (kinematic stage) and then saturates (saturated stage) due to the back reaction of the flow via the Lorentz force induced by the magnetic field \citep[Fig.~1~in][]{Seta19}.
We use the magnetic field from the saturated stage of the simulation with the forcing wavenumber $\kf=5$, Reynolds number $\Re=283$ and magnetic Reynolds number $\Rm=2261$ in a periodic box of non-dimensional size $L=2\pi$ with $512^3$ points \citep{Seta19}. This means that there are $5$ velocity correlation cells along each direction in the domain and a total of $125$ cells in the whole cube. The simulations are for an isothermal gas with incompressible forcing. Thus, the electron number density is roughly uniform within the domain. To consider the effect of thermal electron number density distribution on the rotation measure distribution, we consider the following physically motivated distributions:
\begin{center}
\begin{itemize}
\item a uniform $\ne$ as suggested by isothermal simulations,
\item $\ne$ following the King profile similar to the gas density, i.e., $\ne(r) = \ne(0)(1 + (r/a)^2)^{-3/4}$, where $a$ is the core radius,
\item $\ne$ proportional to $b^{3/2}$ as suggested by the magnetic flux freezing or cooling flow condition (extreme situation, ignores the turbulent nature of the medium).
\end{itemize}
\end{center} 
We assume that the background polarized source density is uniform and the only effect the radiation has while passing through the simulation box is the rotation of its polarization angle, which is quantified by $\RM$. With the fluctuation dynamo generated magnetic field $b$ (normalized to $\brms=1$ over the numerical domain) 
for each $\ne$ (normalized to $\langle \ne \rangle=1$ over the numerical domain) given above, we calculate the $\RM$ at each point within a face of the domain as
\begin{equation} \label{calRM}
\RM (x_i,y_i) = K \sum_{i=1}^{512} \ne(x_i,y_i,z_i) \, b(x_i,y_i,z_i) \, \dd z,
\end{equation}
where $(x_i,y_i,z_i)$ is a coordinate on the grid and $\dd z = 2\pi/512$ is the grid spacing. This involves looking along the sightlines through the core of an elliptical galaxy and gathering statistics from the periodic box. We show the generated $\RM$ maps in \Fig{rm2d} for all three cases: $\ne$ constant (a), $\ne$ King profile (b), and $\ne \propto b^{3/2}$ (c). The shape and size of  structures for a constant $\ne$ ((a) in \Fig{rm2d}) and for $\ne$ following the King profile ((b) in \Fig{rm2d}) are very similar \rev{and} for the King profile case, the intensity of $\RM$ decreases as a function of the distance from the center. For the third case, $\ne \propto b^{3/2}$ ((c) in \Fig{rm2d}), the structures are larger than in the other two cases, with very high $\RM$ values. \rev{On larger scales, the structures look similar for all three cases because of the same magnetic field, but the amplitude and small-scale $\RM$ structures vary depending on the $\ne$ distribution.}

\subsection{Spectra and correlations function of rotation measure maps}
\begin{figure*}
\centering
\includegraphics[width=\columnwidth]{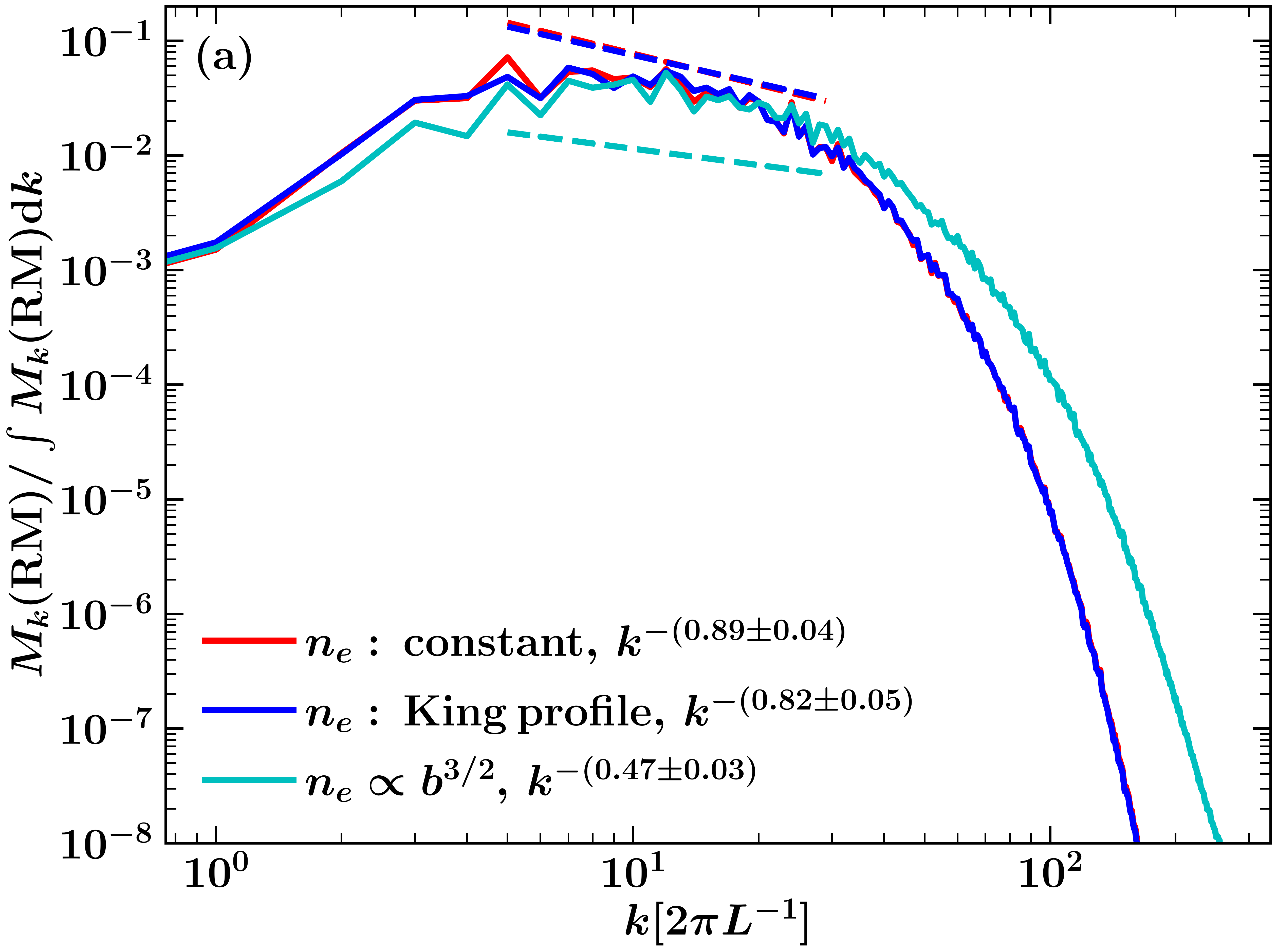}  \hspace{0.25cm}
\includegraphics[width=1.05\columnwidth]{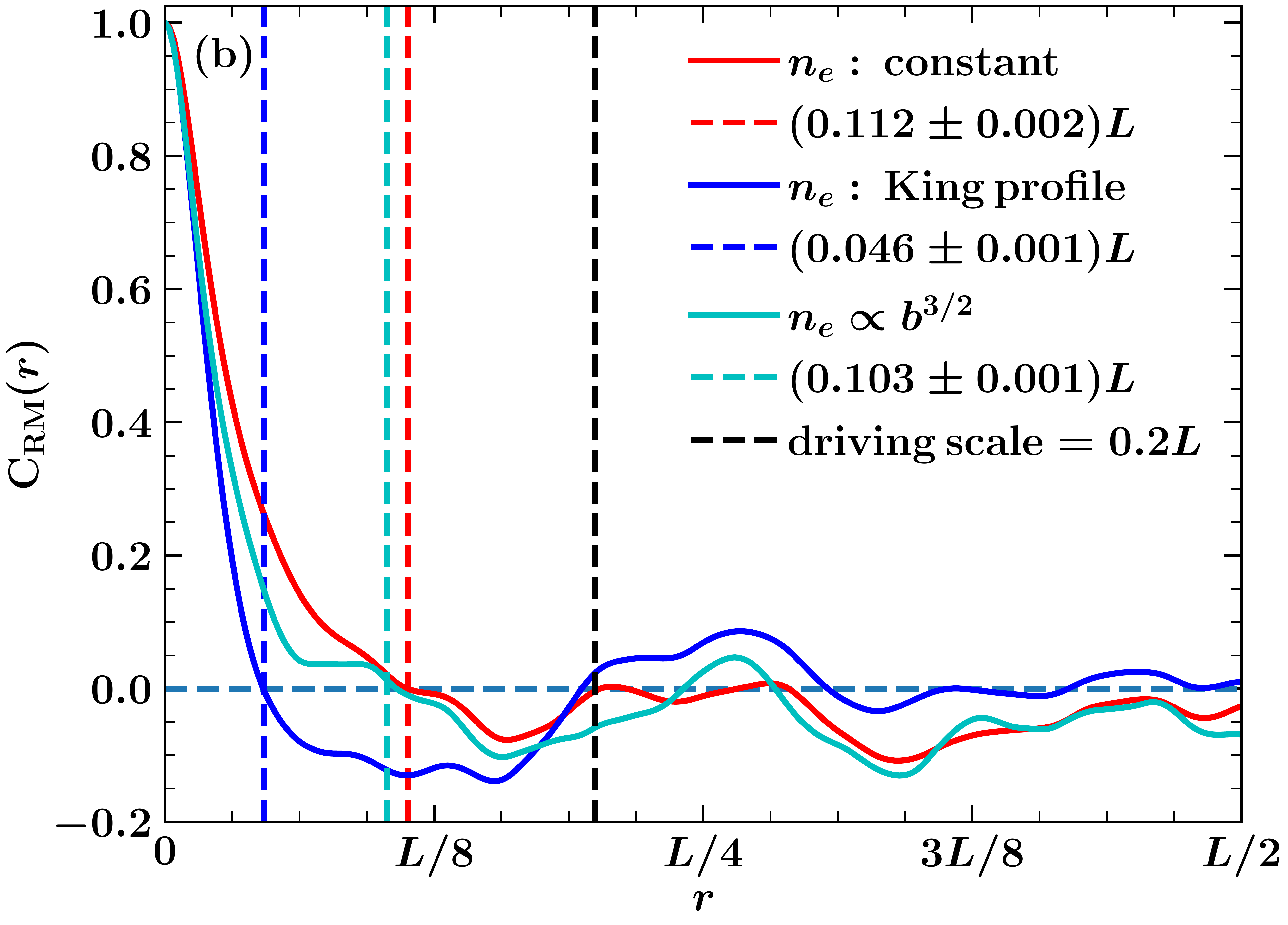}
\caption{One-dimensional (angle-integrated) power spectrum $M_k (\RM)$ (a), and correlation function ${\rm C}_{\RM} (r)$ (b) of the $\RM$ maps shown in \Fig{rm2d}, for the same intermittent magnetic field obtained from a fluctuation dynamo simulation and three different thermal electron density distributions: constant $\ne$ (red), $\ne$ following the King profile (blue), and $\ne \propto b^{3/2}$ (cyan). In the range $3 \le kL/2\pi \le 20$ \citep[in this range the magnetic field roughly follows a $k^{-5/3}$ power-law spectrum; see Fig.~2 of][]{Seta19}, the constant $\ne$ and $\ne$ following the King profile cases have very similar slope, $k^{-0.8}$, whereas for the third case ($\ne \propto b^{3/2}$), the spectrum is flatter, $k^{-0.5}$ \rev{(dashed lines of the same color show the corresponding power-law approximations)}. The right-hand panel shows the correlation function for all three cases, and the dashed vertical lines show the break scale (slope of structure function changes first at this scale, thus referred to as a break scale). This break scale is usually related to the driving scale of the turbulence. However, the driving scale of the turbulence ($0.2 L$) is the same for all three cases, and thus, the difference in break scales is due to the different thermal electron distributions.}
\label{rmcorr}
\end{figure*}

Magnetic fields generated by nonlinear fluctuation dynamo simulations tend to follow a power-law spectrum with slope $3/2$ in the kinematic stage and $-5/3$ in the saturated stage \citep[Fig.~2 in][]{Seta19}. In \Fig{rmcorr}(a), we show the one-dimensional (angle-integrated) power spectrum $M_k$ of $\RM$ using the magnetic field in the saturated stage for the three different electron density distributions discussed above. The slope of the power spectra is similar for the constant $\ne$ and $\ne$ following the King profile case (red and blue lines in \Fig{rmcorr}a), $k^{-0.8}$ and the power spectrum is much flatter for the case with $\ne \propto b^{3/2}$, $M_k\propto k^{-0.5}$. 
Even though all three cases have the same magnetic field, the $\RM$ power spectrum can be different. Then we calculate the second-order structure function ${\rm SF}_{\RM} (r)$ as \citep{Monin_Yaglom1971}:
\begin{equation}
{\rm SF}_{\RM} (r) = \langle |\RM(\vec{x} + \vec{r}) - \RM(\vec{x})|^{2} \rangle,
\end{equation}
where $\vec{x}$ is the position in the two-dimensional $\RM$ plane and $r=|\vec{r}|$ is the length of the displacement vector. Since we have periodic boundary conditions, we consider only half of the periodic domain, i.e., $r$ varies from $0$ to $L/2$.  
The scale at which the slope of the structure function of $\RM$ changes, referred to as the break scale, is usually related to the driving (or outer) scale of the turbulence \citep{Haverkorn08,Anderson15}.  
The structure function of $\RM$ can be used to calculate the $\RM$ correlation function ${\rm C}_{\RM} (r)$  as \citep{Hollins17}:
\begin{equation}
{\rm C}_{\RM} (r) = 1 - \frac{{\rm SF}_{\RM} (r)}{2 \sigma^2},
\end{equation}  
where $2 \sigma^2$ is the value of ${\rm SF}_{\RM}$ at which $\RM$ is no longer correlated, which would also be twice the standard deviation of $\RM$, if there are a sufficient number of correlation cells in the numerical domain.
The break scale can also be determined from the correlation function of $\RM$. It is the length scale at which the correlation function crosses zero for the first time starting from $r=0$. \Fig{rmcorr}(b) shows the correlation function of $\RM$ for different electron density distributions. Even though the turbulence is driven on the same scale ($0.2 L$) for all three cases, the break scale is different. The correlation length of $\RM$ can be calculated by integrating the correlation function as
\begin{equation}
\lb = \int C(r) \, \dd r.
\label{corr}
\end{equation}
Using \Eq{corr}, we calculate the correlation length of $\RM$ for all three cases and find $\lb/L = 0.125 \pm 0.07, 0.080 \pm 0.006$, and $0.052 \pm 0.07$. For this analysis, we see that the correlation length depends not only on the thermal electron density distribution, but is also different from the break scale.

\subsection{Probability distribution function of rotation measure maps}
\begin{figure*}
\centering
\includegraphics[width=\columnwidth]{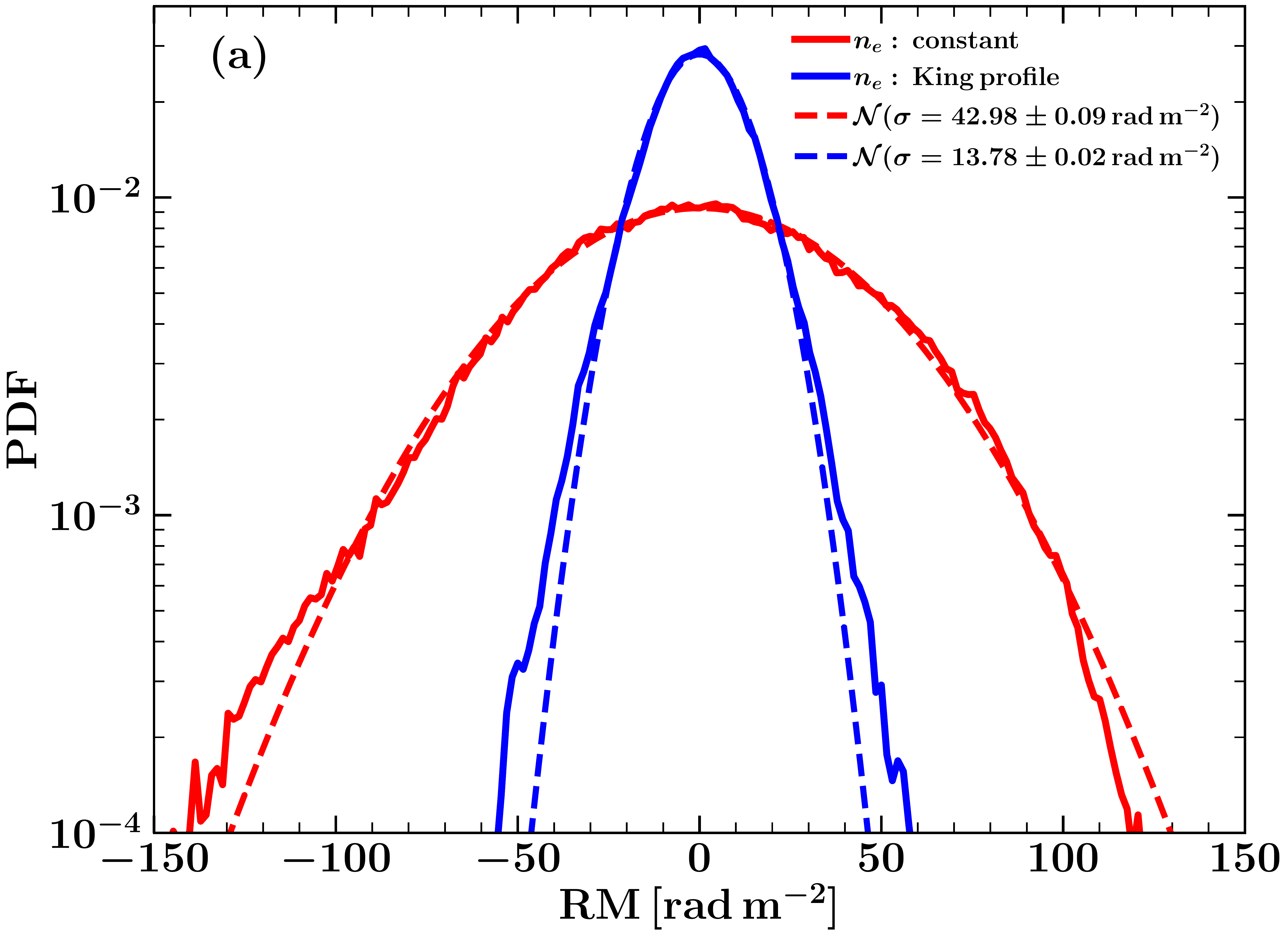}  \hspace{0.25cm}
\includegraphics[width=\columnwidth]{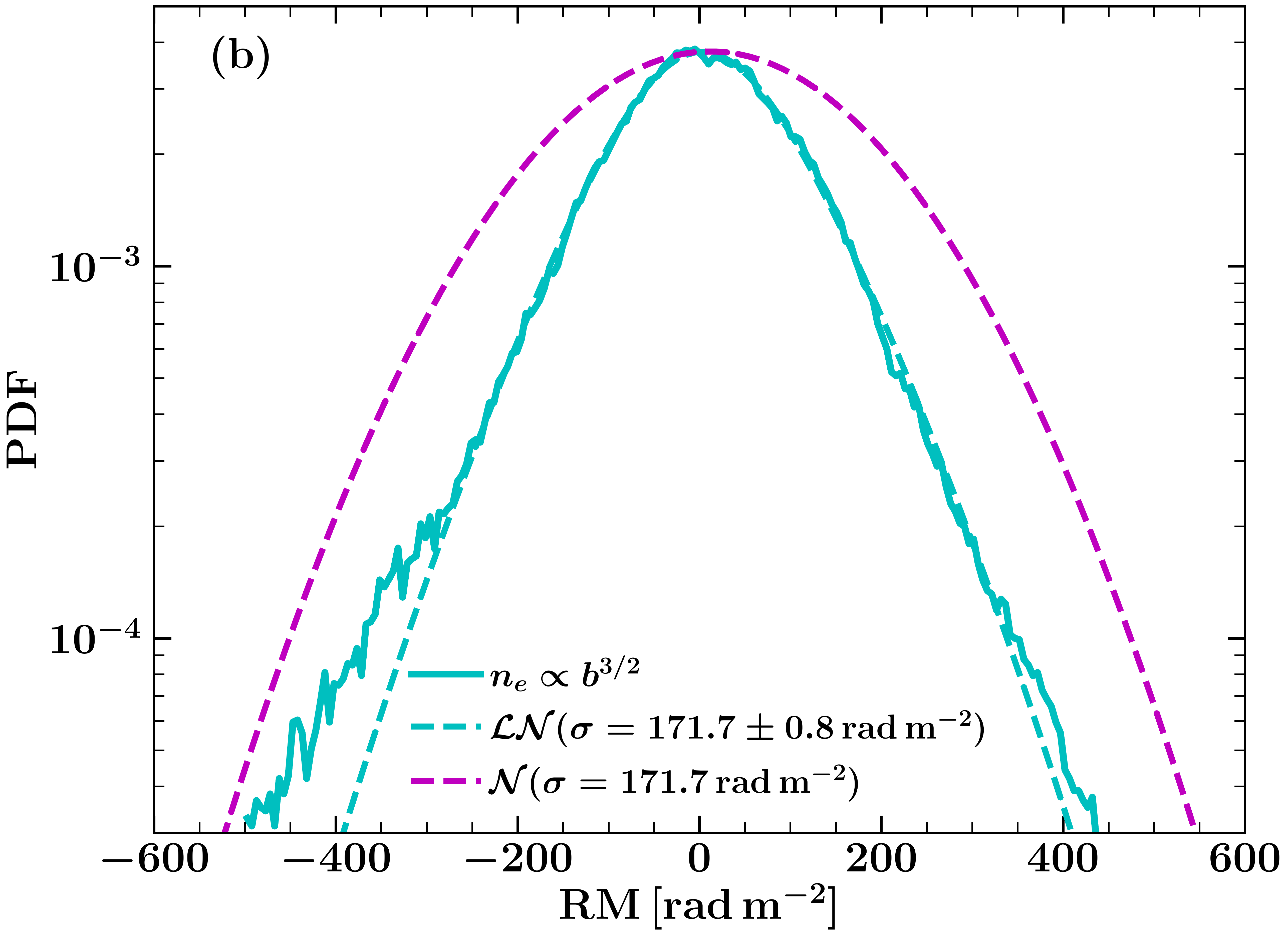}
\caption{(a) PDFs of $\RM$ distribution for the cases where the electron number density $\ne$ is not correlated to the magnetic field strength: $\ne$ constant (red)
and $\ne$ following the King distribution (blue) with $a=L/4=1.25 \ell_0$, where $\ell_0$ is the driving scale of turbulence. Both distributions are roughly Gaussian \rev{(dashed lines of the same color show the corresponding fitted Gaussian distribution)} with mean $\RM \approx 0$ (since a mean magnetic field is absent). The standard deviation $\sigma$ is lower for the King profile. (b) PDF of $\RM$ for the case where $\ne$ is related to magnetic field strength $b$ via the flux freezing or cooling flow condition, i.e., $\ne \propto b^{3/2}$. The distribution (cyan)
is clearly non-Gaussian (the dashed-magenta line shows a Gaussian distribution with the same standard deviation as the fitted distribution) and can be approximated by a product of a Cauchy--Lorentz distribution and a Gaussian function \rev{(the dashed cyan line shows the fitted function)}. For the same magnetic field, the standard deviation and the shape of the $\RM$ distribution differs significantly due to different (physically motivated) thermal electron density distributions.}
\label{rmfluc} 
\end{figure*}

\Fig{rmfluc} shows the probability density function (PDF) of $\RM$ for the three electron number density distributions. In all cases, the mean $\RM\approx0$ and the distribution is symmetric, because there is no large-scale field in the domain. \Fig{rmfluc}(a) shows the PDFs for \rev{two distributions of $\ne$}: uniform and following a King profile with $a=L/4=1.25\ell_0$. Both distributions are \rev{close to a} Gaussian distribution ($\mathcal{N}$) with different standard deviations $\sigma_{\RM}$. The one with the King profile has a smaller standard deviation. The distributions agree well with the analytical expressions, \Eq{sigRM} and \Eq{sigRMne}, respectively. \Fig{rmfluc}(b) shows the $\RM$ distribution when $\ne \propto b^{3/2}$. The distribution is clearly non-Gaussian and can be approximated by a product of a Cauchy--Lorentz distribution and a Gaussian function ($\mathcal{LN}$). The Gaussian or non-Gaussian nature of the $\RM$ distribution in these cases follows from \Eq{defRM} or \Eq{calRM}. In an isotropic random magnetic field, the polarization angle performs a random walk as it rotates randomly along the path length. When $\ne$ does not depend on $b$ directly,  $\RM$ depends on the first power of the magnetic field and the random walk is a Brownian motion which gives rise to a Gaussian distribution.  For the case where $\ne \propto b^{3/2}$, $\RM$ depends on a higher power of the magnetic field and since the underlying magnetic field is intermittent, the resulting $\RM$ distribution is non-Gaussian. In the case of a non-Gaussian distribution, it is difficult to associate a single number (for example $\sigma_{\RM}$) to the distribution but the calculated standard deviation for a non-Gaussian distribution would be far higher than that for a Gaussian distribution (compare the $x$-axis in \Fig{rmfluc}(a) and \Fig{rmfluc}(b)). The calculated standard deviation of the $\RM$ distribution shown in \Fig{rmfluc}(b) is approximately equal to $170 \rad \m^{-2}$, whereas the deviation for both the Gaussian distributions in \Fig{rmfluc}(a) is less than $100 \rad \m^{-2}$. We show that the standard deviation of fluctuations in $\RM$ are significantly different for different thermal electron number density distributions for a given fixed magnetic field. 

\begin{figure}
\centering
\includegraphics[width=\columnwidth]{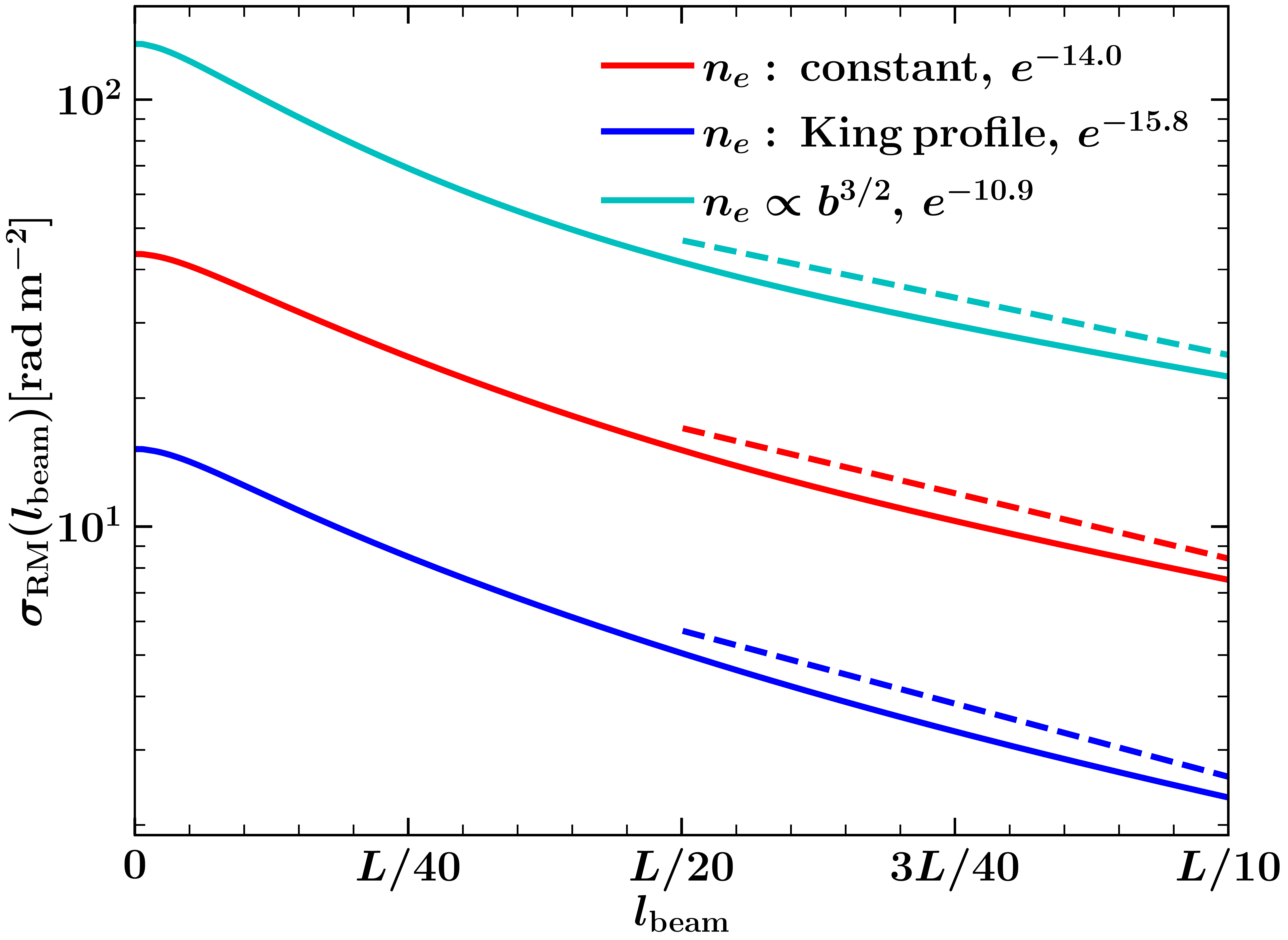}
\caption{Standard deviation of $\RM, \, \sigma_{\RM}$, as a function of beam size $l_{\rm beam}$ for all three electron density distributions. The standard deviation decreases as the beam size increases, and the decrease is exponential for large $l_{\rm beam}$. Even when the beam size is one-tenth of the numerical domain, the standard deviation between the three cases still differs significantly.}
\label{rmlbeam} 
\end{figure}

Furthermore, to account for the effects of beam smearing, we perform a Gaussian smoothing of the $\RM$ maps with a beam size (or smoothing scale) $l_{\rm beam}$ ranging from $0.0$ (no smoothing) to $L/10$ (beam size one-tenth of the numerical domain). The dependence of the standard deviation of $\RM$ on $l_{\rm beam}$ is shown in \Fig{rmlbeam}. For all three cases of the thermal electron density distribution, $\sigma_{\RM}$ decreases (exponentially for high $l_{\rm beam}$), but the difference between the three cases still persists even for strong smoothing of the $\RM$ maps.

Thus, the fluctuations in rotation measure should not be directly associated with the fluctuations in the magnetic fields without considering the fluctuations in thermal electron density.
This sets an expectation for future observations.


\section{Observational Prospects: Magnetic fields using the Laing-Garrington effect}\label{ExiObs}

Observations of magnetic fields in elliptical galaxies are needed to assess the true magnetic field saturation state resulting from fluctuation dynamo action. Several possible approaches are outlined below. Selection effects are discussed in Section~\ref{SelEff}.
\subsection{Rotation Measure Grid} \label{RMG}

The most direct approach would be to measure RM variance using a grid of polarized background radio sources observed through the ISM of a foreground elliptical galaxy. This requires a sufficient number of background sources to enable separation of RM fluctuations arising within and external to the target galaxy. This approach is unlikely to be feasible for any individual galaxy, even if using future highly sensitive radio telescopes, because the sky density of background polarized sources (e.g. \citealt{2014MNRAS.440.3113H} and Fig.~3 from \citealt{Hales2013}) will be too low to yield a sufficient statistical sample behind even an arcmin-sized galaxy \citep[see Section 5.4 in][for an estimate to observe $\RM$s from background extragalactic sources seen through the nearest elliptical galaxy Maffei~1]{Seta2019}. A more promising approach could be to identify elliptical galaxies located in the foreground of extended polarized radio sources (e.g. lobes), through which spatially-resolved RM fluctuations may even be discerned. Depolarization in this scenario would also yield constraints on $\sigma_\textrm{RM}$.

Similarly, polarized fast radio bursts (FRBs) could be used to gather RM statistics. The event rate of FRBs \citep[e.g.][]{Petroff19,Cordes19} is probably insufficient to sample many independent sightlines through any individual foreground galaxy within a realistic timeframe. Furthermore, it may be unrealistic to consider gathering statistics of host galaxy RM contributions to study ellipticals; while the emission mechanism(s) of FRBs and their identification with any particular astronomical object or phenomenon remain unknown, their compelling association with young neutron stars \citep{2019MNRAS.485.4091M} and with environments that scatter radio waves \citep{2019Natur.566..230C} appear to argue against an origin within old stellar populations. Instead, a plausible approach might be to compare a control sample of RMs from FRBs with those located behind elliptical galaxies.

The approaches above could also be attempted without prior knowledge of the foreground elliptical galaxies by searching for correlations with various line tracers. For example, a similar correlation has been identified between polarization properties of background sources and Mg~II absorption line systems, indicating a likely association with magnetic outflows from intervening star forming galaxies \citep{2016ApJ...829..133K}.

\subsection{Laing-Garrington effect} \label{LGsec}

\begin{table*}
\begin{center}
	\caption{List of sources, their redshifts $z$, their projected linear size $\LS$, the depolarization fraction for jets $\DP_{\rm j}$ and counter jets $\DP_{\rm cj}$ with their respective uncertainties \citep{Garr91_1}. 
	The calculated  standard deviation of the rotation measure in jets $\sigma_{\RM_{\rm j}}$, counter jets $\sigma_{\RM_{\rm cj}}$,
	elliptical hosts $\sigma_{\RM_{\rm ellip}}$, and the rms magnetic field in ellipticals with their respective uncertainties for the uniform $\ne$ $\brms(\rm uni)$ and for the $\ne$ following the King profile $\brms({\rm KP}, r=0)$ are also given.}
	\label{tableLG}
	\begin{tabular}{cccccccccc} 
		\hline 
		Name & 
		$z$ & 
		$\LS$ &
		$\DP_{\rm j}$ &
		$\DP_{\rm cj}$ & 
		$\sigma_{\RM_{\rm j}}$ &  
		$\sigma_{\RM_{\rm cj}}$ & 
		$\sigma_{\RM_{\rm ellip}}$ & 
		$\brms(\rm uni)$ & 
		$\brms({\rm KP}, r=0)$\\
		\text{--} & 
		\# & 
		$(\kpc)$ &
		\# & 
		\# &
		$(\rad \m^{-2})$ & 
		$(\rad \m^{-2})$ & 
		$(\rad \m^{-2})$ & 
		$(\mkG\,)$ &
		$(\mkG\,)$ \\
		\hline 
		0017+15 & $2.012$ & $ 88$& $0.61 \pm 0.02$& $0.21 \pm 0.07$& $12.48 \pm 0.41$& $22.17 \pm 2.37$& $18.33 \pm 2.88$& $0.29 \pm 0.05$& $1.82 \pm 0.29$ \\
		0123+32 & $0.794$ & $192$& $0.96 \pm 0.04$& $0.89 \pm 0.03$& $3.59 \pm 1.83$& $6.06 \pm 0.88$& $4.88 \pm 1.73$& $0.15 \pm 0.05$& $1.36 \pm 0.48$ \\
		0225-01 & $2.037$ & $131$& $0.85 \pm 0.03$& $0.08 \pm 0.04$& $7.16 \pm 0.78$& $28.21 \pm 2.79$& $27.29 \pm 2.89$& $0.35 \pm 0.04$& $2.66 \pm 0.28$ \\
		0232-04 & $1.436$ & $110$& $0.81 \pm 0.03$& $0.16 \pm 0.05$& $8.15 \pm 0.72$& $24.03 \pm 2.05$& $22.60 \pm 2.19$& $0.49 \pm 0.05$& $3.42 \pm 0.33$ \\
		0838+13 & $0.684$ & $ 89$& $0.73 \pm 0.04$& $0.25 \pm 0.02$& $9.96 \pm 0.87$& $20.90 \pm 0.60$& $18.37 \pm 0.83$& $0.93 \pm 0.04$& $5.83 \pm 0.26$ \\
		0850+58 & $1.322$ & $129$& $0.57 \pm 0.04$& $0.15 \pm 0.03$& $13.31 \pm 0.83$& $24.45 \pm 1.29$& $20.51 \pm 1.63$& $0.45 \pm 0.04$& $3.42 \pm 0.27$ \\
		1023+06 & $1.699$ & $112$& $0.45 \pm 0.03$& $0.25 \pm 0.06$& $15.86 \pm 0.66$& $20.90 \pm 1.81$& $13.61 \pm 2.88$& $0.24 \pm 0.05$& $1.68 \pm 0.36$ \\
		1115+53 & $1.235$ & $ 77$& $0.86 \pm 0.03$& $0.18 \pm 0.04$& $6.89 \pm 0.80$& $23.24 \pm 1.51$& $22.20 \pm 1.60$& $0.68 \pm 0.05$& $4.00 \pm 0.29$ \\
		1218+33 & $1.519$ & $ 79$& $0.56 \pm 0.02$& $0.07 \pm 0.04$& $13.52 \pm 0.42$& $28.94 \pm 3.11$& $25.60 \pm 3.52$& $0.61 \pm 0.08$& $3.63 \pm 0.50$ \\
		1226+10 & $2.296$ & $ 40$& $0.69 \pm 0.04$& $0.04 \pm 0.02$& $10.81 \pm 0.84$& $31.85 \pm 2.47$& $29.95 \pm 2.65$& $0.59 \pm 0.05$& $2.48 \pm 0.22$ \\
		1241+16 & $0.557$ & $116$& $0.90 \pm 0.06$& $0.41 \pm 0.04$& $5.76 \pm 1.82$& $16.76 \pm 0.92$& $15.74 \pm 1.18$& $0.81 \pm 0.06$& $5.84 \pm 0.44$ \\
		1258+40 & $1.659$ & $172$& $0.84 \pm 0.04$& $0.11 \pm 0.05$& $7.41 \pm 1.01$& $26.37 \pm 2.72$& $25.31 \pm 2.84$& $0.37 \pm 0.04$& $3.22 \pm 0.36$ \\
		1318+11 & $2.171$ & $ 37$& $0.97 \pm 0.04$& $0.06 \pm 0.06$& $3.10 \pm 2.10$& $29.77 \pm 5.29$& $29.61 \pm 5.32$& $0.65 \pm 0.12$& $2.65 \pm 0.48$ \\
		1323+65 & $1.618$ & $ 71$& $0.73 \pm 0.03$& $0.18 \pm 0.03$& $9.96 \pm 0.65$& $23.24 \pm 1.13$& $21.00 \pm 1.29$& $0.49 \pm 0.03$& $2.75 \pm 0.17$ \\
		1634+17 & $1.897$ & $ 55$& $0.77 \pm 0.08$& $0.27 \pm 0.07$& $9.07 \pm 1.80$& $20.31 \pm 2.01$& $18.17 \pm 2.42$& $0.39 \pm 0.05$& $1.95 \pm 0.26$ \\
		1634+58 & $0.985$ & $ 73$& $0.84 \pm 0.10$& $0.25 \pm 0.04$& $7.41 \pm 2.53$& $20.90 \pm 1.21$& $19.54 \pm 1.61$& $0.78 \pm 0.06$& $4.46 \pm 0.37$ \\
		1656+57 & $1.281$ & $ 33$& $0.90 \pm 0.04$& $0.10 \pm 0.10$& $5.76 \pm 1.22$& $26.93 \pm 5.85$& $26.31 \pm 5.99$& $1.19 \pm 0.27$& $4.55 \pm 1.04$ \\
		1709+46 & $0.806$ & $103$& $0.56 \pm 0.02$& $0.15 \pm 0.03$& $13.52 \pm 0.42$& $24.45 \pm 1.29$& $20.37 \pm 1.57$& $0.83 \pm 0.06$& $5.62 \pm 0.43$ \\
		1732+16 & $1.270$ & $132$& $0.68 \pm 0.05$& $0.08 \pm 0.04$& $11.02 \pm 1.05$& $28.21 \pm 2.79$& $25.97 \pm 3.07$& $0.59 \pm 0.07$& $4.53 \pm 0.53$ \\
		1816+47 & $2.225$ & $ 49$& $0.82 \pm 0.06$& $0.06 \pm 0.03$& $7.91 \pm 1.46$& $29.77 \pm 2.65$& $28.70 \pm 2.77$& $0.53 \pm 0.05$& $2.48 \pm 0.24$ \\
		0107+31 & $0.689$ & $364$& $0.65 \pm 0.03$& $0.39 \pm 0.02$& $11.65 \pm 0.62$& $17.22 \pm 0.47$& $12.69 \pm 0.86$& $0.31 \pm 0.02$& $4.00 \pm 0.27$ \\
		0712+53 & $0.064$ & $ 49$& $0.34 \pm 0.03$& $0.28 \pm 0.03$& $18.44 \pm 0.75$& $20.03 \pm 0.84$& $7.82 \pm 2.80$& $1.33 \pm 0.48$& $6.21 \pm 2.22$ \\
		0824+29 & $0.458$ & $129$& $0.86 \pm 0.05$& $0.44 \pm 0.02$& $6.89 \pm 1.33$& $16.08 \pm 0.45$& $14.53 \pm 0.80$& $0.81 \pm 0.04$& $6.15 \pm 0.34$ \\
		0903+16 & $0.411$ & $280$& $0.65 \pm 0.02$& $0.38 \pm 0.03$& $11.65 \pm 0.42$& $17.46 \pm 0.71$& $13.00 \pm 1.03$& $0.53 \pm 0.04$& $5.87 \pm 0.46$ \\
		1001+22 & $0.974$ & $576$& $0.50 \pm 0.04$& $0.36 \pm 0.02$& $14.78 \pm 0.85$& $17.94 \pm 0.49$& $10.17 \pm 1.51$& $0.15 \pm 0.02$& $2.35 \pm 0.35$ \\
		1055+20 & $1.111$ & $399$& $0.88 \pm 0.07$& $0.52 \pm 0.11$& $6.35 \pm 1.97$& $14.35 \pm 2.32$& $12.87 \pm 2.77$& $0.20 \pm 0.04$& $2.60 \pm 0.56$ \\
		1354+19 & $0.720$ & $376$& $0.59 \pm 0.05$& $0.23 \pm 0.03$& $12.89 \pm 1.04$& $21.52 \pm 0.95$& $17.23 \pm 1.42$& $0.41 \pm 0.03$& $5.24 \pm 0.43$ \\
		1548+11 & $1.901$ & $379$& $0.89 \pm 0.13$& $0.33 \pm 0.04$& $6.06 \pm 3.80$& $18.69 \pm 1.02$& $17.68 \pm 1.69$& $0.15 \pm 0.01$& $1.89 \pm 0.18$ \\
		1618+17 & $0.555$ & $430$& $0.84 \pm 0.05$& $0.62 \pm 0.03$& $7.41 \pm 1.27$& $12.27 \pm 0.62$& $9.78 \pm 1.24$& $0.26 \pm 0.03$& $3.64 \pm 0.46$ \\
		1830+28 & $0.594$ & $217$& $0.81 \pm 0.03$& $0.75 \pm 0.04$& $8.15 \pm 0.72$& $9.52 \pm 0.88$& $4.92 \pm 2.08$& $0.18 \pm 0.07$& $1.74 \pm 0.74$ \\
		2325+29 & $1.015$ & $425$& $0.83 \pm 0.05$& $0.60 \pm 0.04$& $7.66 \pm 1.24$& $12.69 \pm 0.83$& $10.11 \pm 1.40$& $0.16 \pm 0.02$& $2.24 \pm 0.31$ \\
		\hline
	\end{tabular}
\end{center}
\end{table*}

\cite{Laing88} and \cite{Garr88} studied radio polarization from Fanaroff--Riley Class II (FRII) sources and observed that the level of fractional polarization was significantly higher on the jet side closer to us than on the counter-jet side pointing away from us. This is referred to as the Laing-Garrington effect. Assuming that both jets have similar intrinsic properties, the difference in fractional polarization can be attributed to depolarization by a foreground screen arising from the X-ray emitting halo of the elliptical galaxy, where the radiation from the more distant side travels more through the screen and is thus more depolarized.

If the elliptical galaxy acts as a Faraday screen, i.e. a medium with magnetic fields and thermal electrons but lacking relativistic electrons (this is a justified assumption since an elliptical galaxy has a scarcity of cosmic ray electrons), and if it is assumed that the jet does not significantly disturb the ISM, we can use historical observations of the Laing-Garrington to probe magnetic fields in elliptical galaxies. We investigate this observational approach below using available historical data.

For a Faraday screen, assuming a Gaussian distribution of Faraday depths, the degree of polarization $p$ at wavelength $\lambda$ is given by \citep{Burn66,Sokoloff1998} 
\footnote{This also assumes no foreground $\RM$ structure; see \cite{Ensslin2003, Murgia2004, Laing2008}, for a discussion on its effect.}
\begin{equation}
p = p_0 \exp(-2 \sigma_{\RM}^2 \lambda^4),
\label{degp}
\end{equation}
where $p_0$ is the maximum degree of polarization and $\sigma_{\RM}$ is the standard deviation of the fluctuations in rotation measure due to the screen.
The depolarization $\DP$ \citep[see Fig.~7 in][for $\DP$ calculated from MHD simulations]{Basu2019} between the wavelengths $\lambda_1$ and $\lambda_2$ with
$\lambda_1 > \lambda_2$ is given by
\begin{equation}
\DP = \exp(-2 \sigma_{\RM}^2 (\lambda_1^4 - \lambda_2^4)).
\label{dp}
\end{equation}
Using \Eq{dp}, we obtain the standard deviation in the rotation measure distribution for the jet $\sigma_{\RM_{\rm j}}$ and counter jet  $\sigma_{\RM_{\rm cj}}$. Then, assuming that the difference is due to the halo of an elliptical galaxy, we derive the standard deviation in the rotation measure distribution for the galaxy $\sigma_{\RM_{\rm ellip}}$ from the standard deviation of rotation measure for the jet and counter jet as follows:
\begin{equation}
\sigma_{\RM_{\rm ellip}} = \sqrt{\sigma_{\RM_{\rm cj}}^2 -\sigma_{\RM_{\rm j}}^2}, 
\end{equation}
Finally, the $\sigma_{\RM_{\rm ellip}}$ is used to estimate the magnetic field strength $\brms$ using the equation \Eq{sigRM} for uniform thermal electron number density and \Eq{sigRMne} for the King profile. 

\begin{figure}
\centering
\includegraphics[width=\columnwidth]{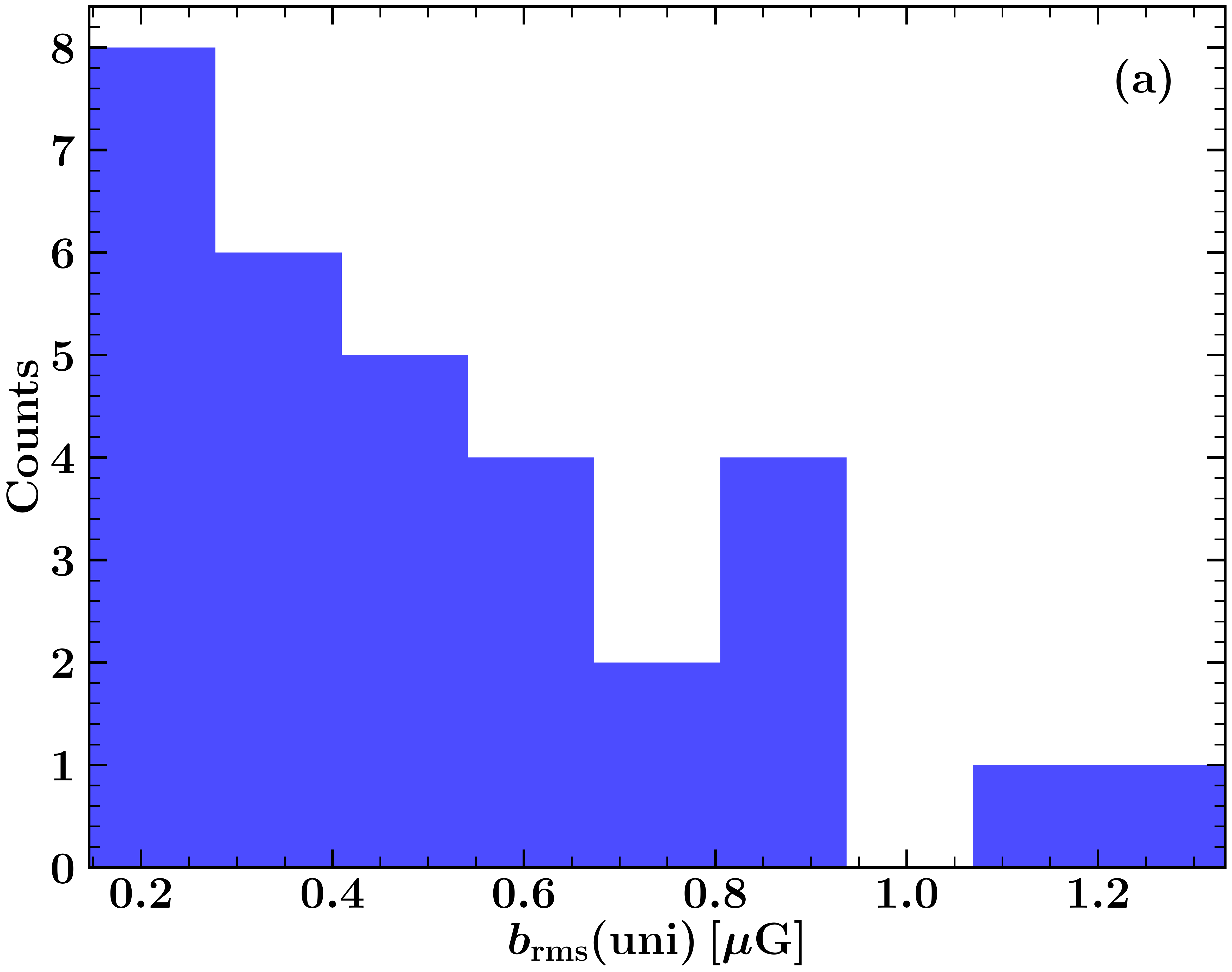} \\ \vspace{0.5cm}
\includegraphics[width=\columnwidth]{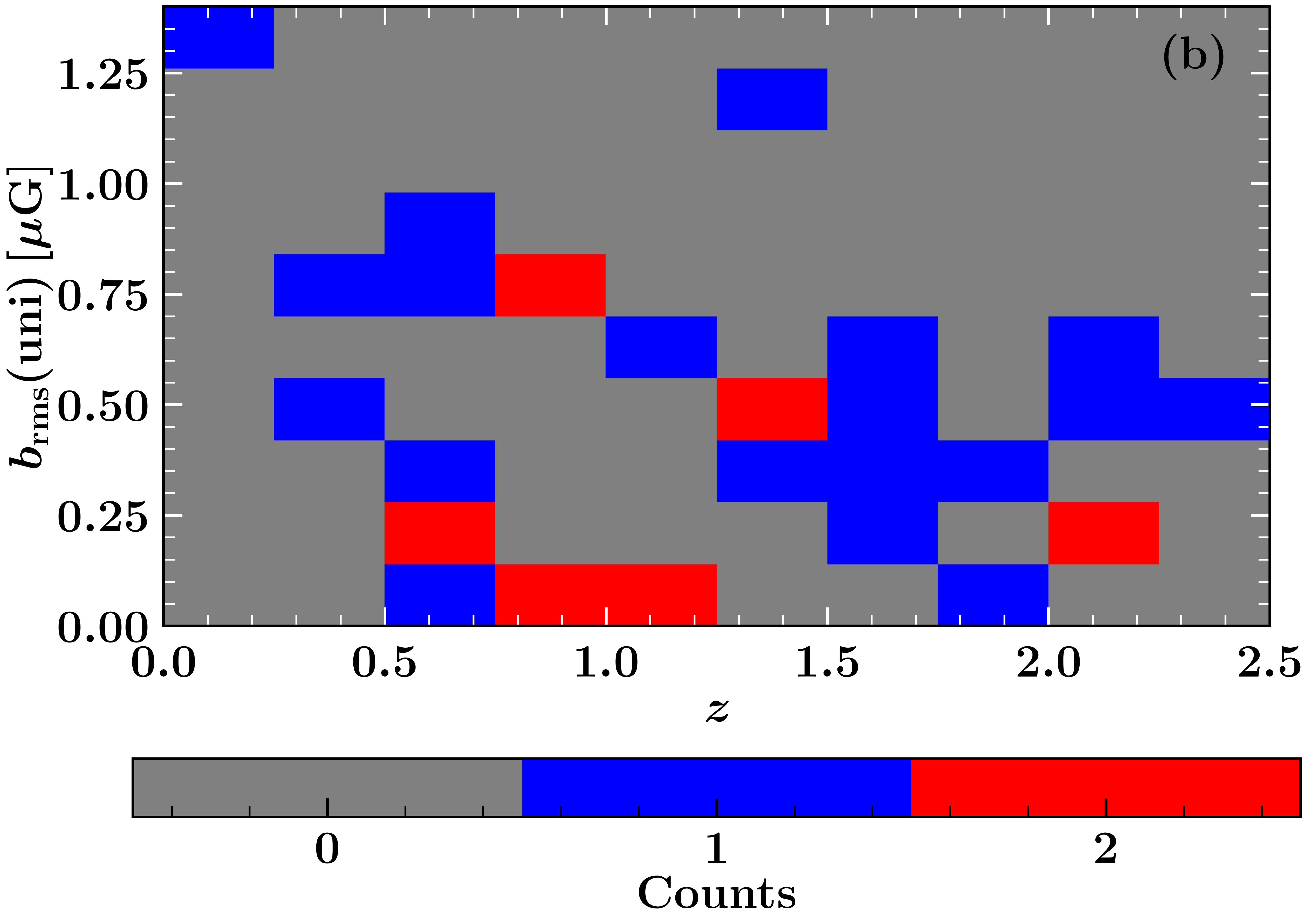}
\caption{(a) Histogram of rms magnetic field strength $\brms(\rm uni) \, [\mkG\,]$ obtained from the asymmetry in polarization properties of jets from sources associated with elliptical galaxies \rev{assuming a uniform thermal electron density $\ne=10^{-2}\cm^{-3}$}. (b) Two dimensional histogram of the redshift of the sources $z$ and $\brms(\rm uni)$. At lower $z$, $\brms$ is somewhat higher compared to that at the higher $z$ values.}
\label{2dbz} 
\end{figure}

\begin{figure}
	\centering
	\includegraphics[width=\columnwidth]{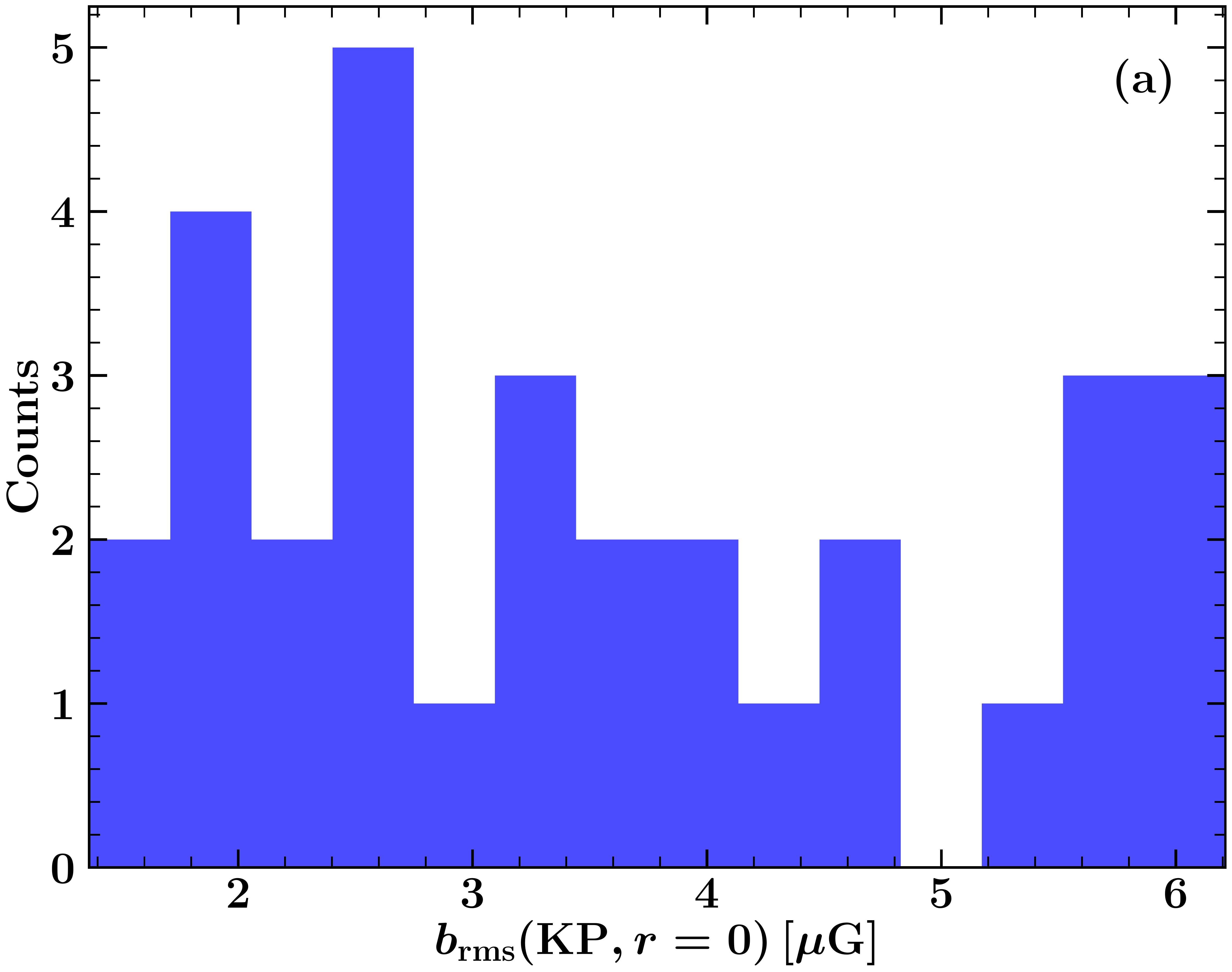} \\ \vspace{0.5cm}
	\includegraphics[width=.71\columnwidth]{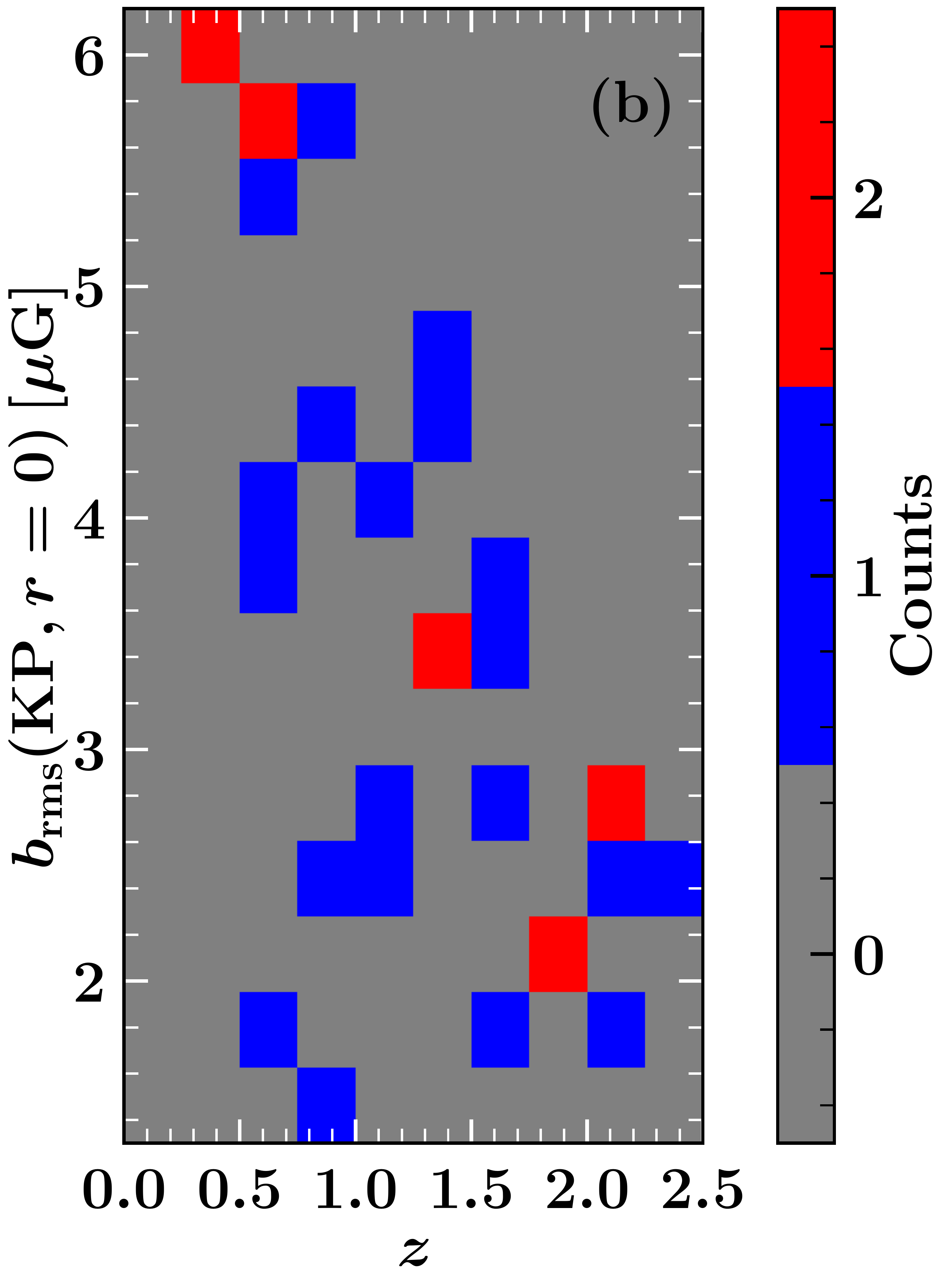}
	\caption{\rev{(a) Histogram of rms magnetic field strength $\brms({\rm KP}, r=0) \, [\mkG\,]$ obtained from the asymmetry in polarization properties of jets from sources associated with elliptical galaxies assuming the King profile for the thermal electron density with the core radius $a=3\kpc$ and the number density at $r=0$ as $\ne(r=0)=10^{-2}\cm^{-3}$.
	(b) Two-dimensional histogram of the redshift of the sources $z$ and $\brms({\rm KP}, r=0)$. At lower $z$, $\brms$ tends to be somewhat (factor 2--3) higher than at higher $z$, although there is significant scatter in the intermediate redshift range, $0.5\leq z \leq 1.75$.}}
\label{2dbzkp} 
\end{figure}	
\cite{Garr91_2} did a similar estimate for a typical single source at low redshift and another source at a high redshift assuming uniform thermal electron density. In Table~\ref{tableLG}, we do this calculation for all the 31 sources studied by \citet{Garr91_1} with varying path lengths ($L \sim$ projected linear size, $\LS$ in Table~\ref{tableLG}) \rev{and two different thermal electron density models (uniform $\ne$ and $\ne$ following the King profile). For the uniform $\ne$ case, we assume $\ne$ to be $10^{-2} \cm^{-3}$ and for the King profile case, we assume $\ne$ at radius $r=0$ to be $10^{-2} \cm^{-3}$ and the core radius $a=3 \kpc$ \citep[see Tab.~4 in][]{Garr91_2}. For both cases, we assume the magnetic field correlation length $\lb=100\pc$ (as expected from the fluctuation dynamo action). The estimated magnetic field for the uniform $\ne$ case $\brms(\rm uni)$ and that at around $r=0$ for the King profile case $\brms({\rm KP}, r=0)$ is given in the last two columns of Table~\ref{tableLG}.
The rms magnetic field for the uniform $\ne$ case ranges from $0.14 \mkG$ to $1.33\mkG$ and for the King profile case ranges from $1.36 \mkG$ to $6.21 \mkG$. The magnetic field strengths in the sample roughly agrees with the estimated value in \Sec{ellip3}.} We propagate the uncertainties reported in the observed values of the depolarization fraction \citep{Garr91_1} to magnetic field strengths. The uncertainties in field strengths ranges from $5\%$ to $42\%$. The uncertainty reported here is due to the measurement error, there can be additional systematics due to variations in $\lb$. \rev{ \Fig{2dbz} (a) and \Fig{2dbzkp} (a)  shows a histogram of magnetic fields strengths in host ellipticals obtained using $\brms(\rm uni)$ and $\brms({\rm KP}, r=0)$ (last two columns of Table~\ref{tableLG}). \Fig{2dbz} (b) and \Fig{2dbzkp} (b) show two-dimensional histograms of the calculated magnetic field versus redshift. From the two-dimensional histograms, it appears that the magnetic field is somewhat stronger in elliptical galaxies at lower redshifts.} However, we caution that the sample size is small and has not been controlled for any selection biases. The typical strengths are broadly consistent with the simulated values shown in \Fig{fig:simulated_brms} (also in agreement with small-scale simulations in \Sec{RMproperties}.), which is an encouraging sign. However, a direct comparison of the redshift trends is not possible, as the observed sample is small and its volume completeness is difficult to judge.

The Laing-Garrington effect can only be employed for elliptical galaxies with an active galactic nucleus (AGN). The jets might contaminate the radio signal as well as the turbulence and magnetic fields in such ellipticals. Thus, it may not be trivial to disentangle the influence of the jet from the magnetic field properties that would arise purely under the influence of supernova driven turbulence in the hot ISM of elliptical galaxies. \rev{In general for ellipticals with an AGN at the center, the galaxy+jet system can give rise to large-scale $\RM$ gradients and asymmetric $\RM$ distributions with non-zero mean $\RM$  \citep[as shown in Fig.~8 of][in comparison to our \Fig{rmfluc}]{Laing2008}. Although the magnetic field strength obtained from the fluctuation dynamo analysis in quiescent ellipticals (\Sec{ellip}) and that using the Laing-Garrington effect for ellipticals with jets are similar, the correlation scale of the magnetic field can be widely different for the two cases.} Moreover, the depolarization asymmetry in jets can also be due to an extended disk of magneto-ionic medium surrounding elliptical galaxies \citep{GN97}, absence of a spherically symmetric halo \citep{Laing2008} or difference in Faraday rotation due to shocked and un-shocked medium \citep{Guidetti12}. To robustly study fluctuation dynamos in elliptical galaxies using the Laing-Garrington effect, a statistically large sample of sources may be required.

\subsection{Synchrotron Emission}

The detection of radio synchrotron radiation could be attempted. We are interested in estimating the synchrotron flux in quiescent elliptical galaxies for which the contribution due to star formation is negligible and that due to the central AGN is absent. Assuming that there is a significant population of relativistic electrons, the flux density from the central region of an elliptical galaxy can be estimated by considering the relationship between minimum energy magnetic field strength and radio luminosity for a spherical radio source. Following \citet{1970ranp.book.....P}, assuming an unresolved source with spectral index $S_\nu \propto \nu^{-0.8}$ over frequencies between $10 \MHz$ to $100 \GHz$, and assuming an ion to electron energy density ratio of 100, the predicted spectral flux density is approximately
\begin{equation}
	12 \,  \textrm{nJy}
        \left(\frac{\brms}{1 \mkG}\right)^{3.5}
        \left(\frac{R}{\textrm{1 kpc}}\right)^{3}
        \left(\frac{\nu}{\textrm{1 GHz}}\right)^{-0.8}
        \left(\frac{D_L}{\textrm{100 Mpc}}\right)^{-2}
        \;,
\end{equation}
where $R$ is the volume radius, $\nu$ is the observing frequency, and $D_L$ is the luminosity distance. Assuming $\nu=100 \MHz$, $R=1 \kpc$ from earlier, $\brms \approx 0.5 \mkG$ from \Sec{ellip3}, and a distance of $50 \Mpc$ ($z\approx0.01$), this equates to approximately $25$~nJy which is unlikely to be feasible with even the next generation of radio telescopes. However, it is plausible that this could be detected by stacking the radio emission over a large sample of suitably selected elliptical galaxies.

In apparent contrast with the flux density estimate above, \cite{Nyland2017} found that cores of early-type galaxies are bright in synchrotron because stars are still being formed in the central regions of the galaxies. They also found a deficiency in radio emission compared to the infrared emission. One possible explanation for this deficiency is the presence of significant magnetic field strengths in the medium\footnote{Another proposed explanation was the quick escape of cosmic ray electrons. However, there is no sign of an outflow or wind in such galaxies. Thus, the cosmic rays transport must be dictated by magnetic fields.}. \cite{Nyland2017} estimated magnetic fields in the range $4\text{--}85\mkG$, with median of $15\mkG$ (an order of magnitude higher than that in the nearby spiral galaxies), but to estimate the magnetic field strength they assumed that those early type galaxies are very similar to present day spiral galaxies. Their assumptions included energy equipartition between cosmic rays and magnetic fields (see \cite{SB2019} for pros and cons of the equipartition assumption), ratio of number density of cosmic ray protons to cosmic ray electrons of $100$, and disc-like or cylindrical geometry of the synchrotron emitting region with gas scale heights similar to a spiral galaxy. Additionally, the emission due to accretion by the central black hole might also contaminate the total synchrotron intensity. Based on our simple estimates from \Sec{ellip3} and the indirect observational result from Section~\ref{LGsec}, we argue that magnetic field strengths in elliptical galaxies are probably an order of magnitude lower than that in spirals and certainly not higher.

\subsection{Gravitational Lensing}

Strong gravitational lens systems could be used to measure differences in RM or depolarization between multiple (and/or possibly spatially resolved) images of a background polarized radio source that have been lensed by a foreground elliptical galaxy \citep{GRB85,1999MNRAS.307L...1P,Narasimha2004}\footnote{Such an approach is shown to be successful for spiral galaxies in \cite{Mao17}.}. A mass model for the lensing system is required. A complication with this approach is that the local group or cluster environment will also contribute toward $\sigma_\textrm{RM}$.

\cite{GRB85} studied the radio polarization signal from two gravitationally lensed images of the quasar 0957+561 ($z\approx1.41$) at multiple wavelengths and found that the rotation measures of the two images differ by $100 \rad \m^{-2}$. This difference can be attributed to the magnetic field in the ISM of the lensing (cD) galaxy at $z\approx0.36$. So, to estimate the magnetic field in the galaxy using \Eq{defRM} for the quasar at a redshift $z$, we can write $\RM$ in terms of mean quantities as
\begin{equation}
\RM \approx \left(0.81 \frac{\langle \ne \rangle}{\cm^{-3}} \frac{\langle b \rangle}{\mkG} \frac{L}{\pc} \right)\bigg/\left(1+z\right)^2,
\label{grbb}
\end{equation}
where $\langle \cdots \rangle$ denotes the average over the path length. Assuming $\langle \ne \rangle \approx 0.01 \cm^{-3}$ and $L\approx30 \kpc$, for $\RM=100 \rad \m^{-2}$ and 
$z=1.41$, we obtain  $\langle b \rangle\simeq2.5\mkG$. This is comparable in strength to magnetic fields observed in the Milky Way and nearby spiral galaxies \citep{Fletcher10,Haverkorn15,Beck2016}. Also, since the $\langle b \rangle$ is not close to zero, it refers to the presence of a large-scale or ordered field. Furthermore, such strong ordered fields are not expected in cD galaxies, and thus, the $\RM$ difference cannot be completely associated with the galaxy. The non-zero strong ordered field could be due to the contribution from the cluster's magnetic field, which even if random \citep[on scales of $10 \kpc$,][]{SSH06}, can act as a coherent field over the scale of the galaxy. This argument is supported by the mass modelling of the system, which suggests that the lens must also have some contribution from the cluster to explain the lensing observations \citep{GRB85}. Thus, it then becomes difficult to differentiate between the contribution to the $\RM$ difference due to the galaxy's and cluster's magnetic fields. 

Alternatively, rather than examining RMs directly, RM variance could be probed by observing depolarization similar to Section~\ref{LGsec} but over the spatial extent of the foreground lensing elliptical galaxy within an extended background radio source \citep{1988A&A...194...79S,Laing88,Garr88}. More recent attempts to statistically measure $\RM$ excess due to intervening galaxies \citep{FarnesEA2014, MalikCS2020, LanP2020} can also be extended to study magnetic fields in elliptical galaxies by considering the morphology of the intervening galaxies in the statistical sample.

\subsection{Selection Effects}\label{SelEff}

Observations must carefully consider selection effects so as to minimize potential bias from coherent magnetic fields and from unrelated RM fluctuations. The observations should ideally target giant (core) ellipticals that as a class exhibit negligible rotation \citep{2009ApJS..182..216K}. These galaxies are typically located within the central regions of clusters, so it will be important to discriminate between RM fluctuations arising in the interstellar and intracluster gas \citep{2006MNRAS.373..469B,2018MNRAS.474..547K}. Selection should also ideally target ellipticals in hydrostatic equilibrium \citep{St12,2017MNRAS.468.3883P}. Note that a simple selection on red galaxies could lead to misleading results \citep[e.g.][]{2018MNRAS.476.5284E}. Evolution with redshift in the source sample \citep{2010ApJ...709.1018V,2017MNRAS.468.3883P} will need to be taken into account when compiling statistics on saturation RM variance.

Some of the approaches described in the sections above could be tested now using a combination of data from surveys such as the SDSS \citep{2000AJ....120.1579Y} and NVSS \citep{1998AJ....115.1693C}, or in the near future, for example using forthcoming data from the VLA Sky Survey \citep{2019arXiv190701981L} and eventually from the Square Kilometre Array \citep{Taylor2015}.


\section{Conclusions} \label{conc}

We propose that observations of magnetic fields in {\it quiescent} elliptical galaxies would provide a direct confirmation of the operation of a fluctuation dynamo. We demonstrate that the turbulence driven by Type Ia supernova explosions in elliptical galaxies would generate magnetic fields of a sufficient strength ($0.2~\text{--}~1 \mkG$) to be observable. We examine observational prospects for measuring the magnetic saturation state of fluctuation dynamos using elliptical galaxies. We take the first steps toward characterizing the properties of magnetic fields in elliptical galaxies and analyzing historical observations of the Laing-Garrington effect to estimate magnetic field strengths in a small sample of host galaxies (\Tab{tableLG}). We obtain magnetic field strengths in those host galaxies lies in the range $0.14~\text{--}~1.33 \mkG$ \rev{if a uniform thermal electron density is assumed and $1.36~\text{--}~6.21 \mkG$ if the King profile for the thermal electron density is assumed}. We also examine the redshift dependence of the observed magnetic field strengths (\Fig{2dbz}(b) and \Fig{2dbzkp}(b)) and \rev{find that on average, the magnetic field increases by a factor of 2--3 as the redshift decreases}, but it is difficult to draw robust conclusions from the small sample. More generally, we also show that fluctuations in rotation measures from background polarized sources must not directly be associated with the fluctuations in the magnetic fields as that holds true only for uniform or very weakly varying thermal electron number density.

If the proposed observations \rev{of magnetic fields in elliptical galaxies} reveal much lower saturation magnetic field strengths than our estimate, this would imply gaps in our theoretical understanding of fluctuation dynamos, and would also imply that the mean-field dynamo in spiral galaxies are probably seeded with a weaker magnetic field. This in turn would require a much more efficient mean-field dynamo mechanism than presently known to explain the observed magnetic fields in \rev{the Milky Way and} nearby spiral galaxies.

\acknowledgments
We thank the anonymous referee for a thorough review of this work. AS thanks Andrew Fletcher, Sui Ann Mao, Anvar Shukurov, Kandaswamy Subramanian, and Rainer Beck for very useful discussions. AS thanks Govind Nandakumar for a useful discussion on error analysis.  We thank Sharanya Sur for providing valuable comments on the paper. 
LFSR acknowledges funding from the European Research Council (ERC) under the European Union's Horizon 2020 research and innovation programme (grant agreement No 772663).
CF acknowledges funding provided by the Australian Research Council (Discovery Project DP170100603 and Future Fellowship FT180100495), and the Australia-Germany Joint Research Cooperation Scheme (UA-DAAD). CAH acknowledges support from the European Union's Horizon 2020 research and innovation programme under the Marie Sk{\l}odowska-Curie grant agreement No 705332. We further acknowledge high-performance computing resources provided by the Leibniz Rechenzentrum and the Gauss Centre for Supercomputing (grants~pr32lo, pr48pi and GCS Large-scale project~10391), the Australian National Computational Infrastructure (grant~ek9) in the framework of the National Computational Merit Allocation Scheme and the ANU Allocation Scheme.

\bibliographystyle{aasjournal}

\begin{thebibliography}{}
\expandafter\ifx\csname natexlab\endcsname\relax\def\natexlab#1{#1}\fi
\providecommand{\url}[1]{\href{#1}{#1}}

\bibitem[{{Anderson} {et~al.}(2015){Anderson}, {Gaensler}, {Feain}, \&
  {Franzen}}]{Anderson15}
{Anderson}, C.~S., {Gaensler}, B.~M., {Feain}, I.~J., \& {Franzen}, T.~M.~O.
  2015, \apj, 815, 49

\bibitem[{{Arshakian} {et~al.}(2009){Arshakian}, {Beck}, {Krause}, \&
  {Sokoloff}}]{Arshakian09}
{Arshakian}, T.~G., {Beck}, R., {Krause}, M., \& {Sokoloff}, D. 2009, \aap,
  494, 21

\bibitem[{{Baldry} {et~al.}(2006){Baldry}, {Balogh}, {Bower}, {Glazebrook},
  {Nichol}, {Bamford}, \& {Budavari}}]{2006MNRAS.373..469B}
{Baldry}, I.~K., {Balogh}, M.~L., {Bower}, R.~G., {et~al.} 2006, \mnras, 373,
  469

\bibitem[{{Basu} {et~al.}(2019){Basu}, {Fletcher}, {Mao}, {Burkhart}, {Beck},
  \& {Schnitzeler}}]{Basu2019}
{Basu}, A., {Fletcher}, A., {Mao}, S.~A., {et~al.} 2019, Galaxies, 7, 89

\bibitem[{{Beck}(2016)}]{Beck2016}
{Beck}, R. 2016, \araa, 24, 4

\bibitem[{{Beck} {et~al.}(1996){Beck}, {Brandenburg}, {Moss}, {Shukurov}, \&
  {Sokoloff}}]{Beck1996}
{Beck}, R., {Brandenburg}, A., {Moss}, D., {Shukurov}, A., \& {Sokoloff}, D.
  1996, \araa, 34, 155

\bibitem[{{Benson} \& {Devereux}(2010)}]{Benson2010}
{Benson}, A.~J., \& {Devereux}, N. 2010, \mnras, 402, 2321

\bibitem[{{Bhat} \& {Subramanian}(2013)}]{BS13}
{Bhat}, P., \& {Subramanian}, K. 2013, \mnras, 429, 2469

\bibitem[{{Brandenburg} \& {Subramanian}(2005)}]{BS2005}
{Brandenburg}, A., \& {Subramanian}, K. 2005, \physrep, 417, 1

\bibitem[{{Bregman} {et~al.}(2005){Bregman}, {Miller}, {Athey}, \&
  {Irwin}}]{Bregman05}
{Bregman}, J.~N., {Miller}, E.~D., {Athey}, A.~E., \& {Irwin}, J.~A. 2005,
  \apj, 635, 1031

\bibitem[{{Bregman} {et~al.}(2001){Bregman}, {Miller}, \& {Irwin}}]{Bregman01}
{Bregman}, J.~N., {Miller}, E.~D., \& {Irwin}, J.~A. 2001, \apjl, 553, L125

\bibitem[{{Burn}(1966)}]{Burn66}
{Burn}, B.~J. 1966, \mnras, 133, 67

\bibitem[{{Cappellari} {et~al.}(2011){Cappellari}, {Emsellem}, {Krajnovi{\'c}},
  {McDermid}, {Serra}, {Alatalo}, {Blitz}, {Bois}, {Bournaud}, {Bureau},
  {Davies}, {Davis}, {de Zeeuw}, {Khochfar}, {Kuntschner}, {Lablanche},
  {Morganti}, {Naab}, {Oosterloo}, {Sarzi}, {Scott}, {Weijmans}, \&
  {Young}}]{Atlas3d3}
{Cappellari}, M., {Emsellem}, E., {Krajnovi{\'c}}, D., {et~al.} 2011, \mnras,
  416, 1680

\bibitem[{{Cappellaro} {et~al.}(1999){Cappellaro}, {Evans}, \&
  {Turatto}}]{capp99}
{Cappellaro}, E., {Evans}, R., \& {Turatto}, M. 1999, \aap, 351, 459

\bibitem[{{CHIME/FRB Collaboration} {et~al.}(2019){CHIME/FRB Collaboration},
  {Amiri}, {Bandura}, {Bhardwaj}, {Boubel}, {Boyce}, {Boyle}, {Brar},
  {Burhanpurkar}, {Chawla}, {Cliche}, {Cubranic}, {Deng}, {Denman}, {Dobbs},
  {Fandino}, {Fonseca}, {Gaensler}, {Gilbert}, {Giri}, {Good}, {Halpern},
  {Hanna}, {Hill}, {Hinshaw}, {H{\"o}fer}, {Josephy}, {Kaspi}, {Landecker},
  {Lang}, {Masui}, {Mckinven}, {Mena-Parra}, {Merryfield}, {Milutinovic},
  {Moatti}, {Naidu}, {Newburgh}, {Ng}, {Patel}, {Pen}, {Pinsonneault-Marotte},
  {Pleunis}, {Rafiei-Ravandi}, {Ransom}, {Renard}, {Scholz}, {Shaw}, {Siegel},
  {Smith}, {Stairs}, {Tendulkar}, {Tretyakov}, {Vanderlinde}, \&
  {Yadav}}]{2019Natur.566..230C}
{CHIME/FRB Collaboration}, {Amiri}, M., {Bandura}, K., {et~al.} 2019, \nat,
  566, 230

\bibitem[{{Condon} {et~al.}(1998){Condon}, {Cotton}, {Greisen}, {Yin},
  {Perley}, {Taylor}, \& {Broderick}}]{1998AJ....115.1693C}
{Condon}, J.~J., {Cotton}, W.~D., {Greisen}, E.~W., {et~al.} 1998, \aj, 115,
  1693

\bibitem[{{Cordes} \& {Chatterjee}(2019)}]{Cordes19}
{Cordes}, J.~M., \& {Chatterjee}, S. 2019, \araa, 57, 417

\bibitem[{{de Plaa} {et~al.}(2012){de Plaa}, {Zhuravleva}, {Werner}, {Kaastra},
  {Churazov}, {Smith}, {Raassen}, \& {Grange}}]{PZ2012}
{de Plaa}, J., {Zhuravleva}, I., {Werner}, N., {et~al.} 2012, \aap, 539, A34

\bibitem[{{Dyson} \& {Williams}(1997)}]{DW97}
{Dyson}, J.~E., \& {Williams}, D.~A. 1997, {The physics of the interstellar
  medium}, 2nd edn. (CRC Press), doi:10.1201/9780585368115

\bibitem[{{Emsellem} {et~al.}(2011){Emsellem}, {Cappellari}, {Krajnovi{\'c}},
  {Alatalo}, {Blitz}, {Bois}, {Bournaud}, {Bureau}, {Davies}, {Davis}, {de
  Zeeuw}, {Khochfar}, {Kuntschner}, {Lablanche}, {McDermid}, {Morganti},
  {Naab}, {Oosterloo}, {Sarzi}, {Scott}, {Serra}, {van de Ven}, {Weijmans}, \&
  {Young}}]{Atlas3d2}
{Emsellem}, E., {Cappellari}, M., {Krajnovi{\'c}}, D., {et~al.} 2011, \mnras,
  414, 888

\bibitem[{{En{\ss}lin} \& {Vogt}(2003)}]{Ensslin2003}
{En{\ss}lin}, T.~A., \& {Vogt}, C. 2003, \aap, 401, 835

\bibitem[{{Evans} {et~al.}(2018){Evans}, {Parker}, \&
  {Roberts}}]{2018MNRAS.476.5284E}
{Evans}, F.~A., {Parker}, L.~C., \& {Roberts}, I.~D. 2018, \mnras, 476, 5284

\bibitem[{{Fabian}(1994)}]{Fabian94}
{Fabian}, A.~C. 1994, Annual Review of Astronomy and Astrophysics, 32, 277

\bibitem[{{Fabian} {et~al.}(2005){Fabian}, {Reynolds}, {Taylor}, \&
  {Dunn}}]{Fabian2005}
{Fabian}, A.~C., {Reynolds}, C.~S., {Taylor}, G.~B., \& {Dunn}, R.~J.~H. 2005,
  \mnras, 363, 891

\bibitem[{{Farnes} {et~al.}(2014){Farnes}, {O'Sullivan}, {Corrigan}, \&
  {Gaensler}}]{FarnesEA2014}
{Farnes}, J.~S., {O'Sullivan}, S.~P., {Corrigan}, M.~E., \& {Gaensler}, B.~M.
  2014, \apj, 795, 63

\bibitem[{{Federrath}(2016)}]{Fed16}
{Federrath}, C. 2016, Journal of Plasma Physics, 82, 535820601

\bibitem[{{Federrath} {et~al.}(2011){Federrath}, {Chabrier}, {Schober},
  {Banerjee}, {Klessen}, \& {Schleicher}}]{Fed11}
{Federrath}, C., {Chabrier}, G., {Schober}, J., {et~al.} 2011, \prl, 107,
  114504

\bibitem[{{Federrath} {et~al.}(2009){Federrath}, {Klessen}, \&
  {Schmidt}}]{FederrathKlessenSchmidt2009}
{Federrath}, C., {Klessen}, R.~S., \& {Schmidt}, W. 2009, \apj, 692, 364

\bibitem[{{Federrath} {et~al.}(2014){Federrath}, {Schober}, {Bovino}, \&
  {Schleicher}}]{Fed14}
{Federrath}, C., {Schober}, J., {Bovino}, S., \& {Schleicher}, D. R.~G. 2014,
  \apj, 797, L19

\bibitem[{{Felten}(1996)}]{Felten96}
{Felten}, J.~E. 1996, Astronomical Society of the Pacific Conference Series,
  Vol.~88, {Mitigating the Baryon Crisis in Clusters: Can Magnetic Pressure be
  Important?}, ed. V.~{Trimble} \& A.~{Reisenegger}, 271

\bibitem[{{Fletcher}(2010)}]{Fletcher10}
{Fletcher}, A. 2010, in The Dynamic Interstellar Medium: A Celebration of the
  Canadian Galactic Plane Survey, ed. R.~{Kothes}, T.~L. {Landecker}, \& A.~G.
  {Willis}, Vol. 438, 197

\bibitem[{{Fletcher} {et~al.}(2011){Fletcher}, {Beck}, {Shukurov},
  {Berkhuijsen}, \& {Horellou}}]{Fletcher2011}
{Fletcher}, A., {Beck}, R., {Shukurov}, A., {Berkhuijsen}, E.~M., \&
  {Horellou}, C. 2011, \mnras, 412, 2396

\bibitem[{{Forman} {et~al.}(1985){Forman}, {Jones}, \& {Tucker}}]{Forman85}
{Forman}, W., {Jones}, C., \& {Tucker}, W. 1985, \apj, 293, 102

\bibitem[{{Gaensler} {et~al.}(2005){Gaensler}, {Haverkorn}, {Staveley-Smith},
  {Dickey}, {McClure-Griffiths}, {Dickel}, \& {Wolleben}}]{Gaenslar2005}
{Gaensler}, B.~M., {Haverkorn}, M., {Staveley-Smith}, L., {et~al.} 2005,
  Science, 307, 1610

\bibitem[{{Garrington} \& {Conway}(1991)}]{Garr91_2}
{Garrington}, S.~T., \& {Conway}, R.~G. 1991, \mnras, 250, 198

\bibitem[{{Garrington} {et~al.}(1991){Garrington}, {Conway}, \&
  {Leahy}}]{Garr91_1}
{Garrington}, S.~T., {Conway}, R.~G., \& {Leahy}, J.~P. 1991, \mnras, 250, 171

\bibitem[{{Garrington} {et~al.}(1988){Garrington}, {Leahy}, {Conway}, \&
  {Laing}}]{Garr88}
{Garrington}, S.~T., {Leahy}, J.~P., {Conway}, R.~G., \& {Laing}, R.~A. 1988,
  \nat, 331, 147

\bibitem[{{Gopal-Krishna} \& {Nath}(1997)}]{GN97}
{Gopal-Krishna}, \& {Nath}, B.~B. 1997, \aap, 326, 45

\bibitem[{{Greenfield} {et~al.}(1985){Greenfield}, {Roberts}, \&
  {Burke}}]{GRB85}
{Greenfield}, P.~D., {Roberts}, D.~H., \& {Burke}, B.~F. 1985, \apj, 293, 370

\bibitem[{{Guidetti} {et~al.}(2012){Guidetti}, {Laing}, {Croston}, {Bridle}, \&
  {Parma}}]{Guidetti12}
{Guidetti}, D., {Laing}, R.~A., {Croston}, J.~H., {Bridle}, A.~H., \& {Parma},
  P. 2012, \mnras, 423, 1335

\bibitem[{{Hales}(2013)}]{Hales2013}
{Hales}, C.~A. 2013, ArXiv e-prints, arXiv:1312.4602

\bibitem[{{Hales} {et~al.}(2014){Hales}, {Norris}, {Gaensler}, \&
  {Middelberg}}]{2014MNRAS.440.3113H}
{Hales}, C.~A., {Norris}, R.~P., {Gaensler}, B.~M., \& {Middelberg}, E. 2014,
  \mnras, 440, 3113

\bibitem[{{Haugen} {et~al.}(2004){Haugen}, {Brandenburg}, \&
  {Dobler}}]{Haugen2004}
{Haugen}, N.~E., {Brandenburg}, A., \& {Dobler}, W. 2004, \pre, 70, 016308

\bibitem[{{Haverkorn}(2015)}]{Haverkorn15}
{Haverkorn}, M. 2015, in Astrophysics and Space Science Library, Vol. 407,
  Magnetic Fields in Diffuse Media, ed. A.~{Lazarian}, E.~M. {de Gouveia Dal
  Pino}, \& C.~{Melioli}, 483

\bibitem[{{Haverkorn} {et~al.}(2008){Haverkorn}, {Brown}, {Gaensler}, \&
  {McClure-Griffiths}}]{Haverkorn08}
{Haverkorn}, M., {Brown}, J.~C., {Gaensler}, B.~M., \& {McClure-Griffiths},
  N.~M. 2008, \apj, 680, 362

\bibitem[{{Hernquist}(1990)}]{Hernquist}
{Hernquist}, L. 1990, \apj, 356, 359

\bibitem[{{Hollins} {et~al.}(2017){Hollins}, {Sarson}, {Shukurov}, {Fletcher},
  \& {Gent}}]{Hollins17}
{Hollins}, J.~F., {Sarson}, G.~R., {Shukurov}, A., {Fletcher}, A., \& {Gent},
  F.~A. 2017, \apj, 850, 4

\bibitem[{{Houde} {et~al.}(2013){Houde}, {Fletcher}, {Beck}, {Hildebrand},
  {Vaillancourt}, \& {Stil}}]{Houde2013}
{Houde}, M., {Fletcher}, A., {Beck}, R., {et~al.} 2013, \apj, 766, 49

\bibitem[{{Iacobelli} {et~al.}(2013){Iacobelli}, {Haverkorn}, {Orr{\'u}},
  {Pizzo}, {Anderson}, {Beck}, {Bell}, {Bonafede}, {Chyzy}, {Dettmar},
  {En{\ss}lin}, {Heald}, {Horellou}, {Horneffer}, {Jurusik}, {Junklewitz},
  {Kuniyoshi}, {Mulcahy}, {Paladino}, {Reich}, {Scaife}, {Sobey},
  {Sotomayor-Beltran}, {Alexov}, {Asgekar}, {Avruch}, {Bell}, {van Bemmel},
  {Bentum}, {Bernardi}, {Best}, {B{\i}rzan}, {Breitling}, {Broderick}, {Brouw},
  {Br{\"u}ggen}, {Butcher}, {Ciardi}, {Conway}, {de Gasperin}, {de Geus},
  {Duscha}, {Eisl{\"o}ffel}, {Engels}, {Falcke}, {Fallows}, {Ferrari},
  {Frieswijk}, {Garrett}, {Grie{\ss}meier}, {Gunst}, {Hamaker}, {Hassall},
  {Hessels}, {Hoeft}, {H{\"o}randel}, {Jelic}, {Karastergiou}, {Kondratiev},
  {Koopmans}, {Kramer}, {Kuper}, {van Leeuwen}, {Macario}, {Mann}, {McKean},
  {Munk}, {Pandey-Pommier}, {Polatidis}, {R{\"o}ttgering}, {Schwarz}, {Sluman},
  {Smirnov}, {Stappers}, {Steinmetz}, {Tagger}, {Tang}, {Tasse}, {Toribio},
  {Vermeulen}, {Vocks}, {Vogt}, {van Weeren}, {Wise}, {Wucknitz}, {Yatawatta},
  {Zarka}, \& {Zensus}}]{Iacobelli2013}
{Iacobelli}, M., {Haverkorn}, M., {Orr{\'u}}, E., {et~al.} 2013, \aap, 558, A72

\bibitem[{{Kim} {et~al.}(2016){Kim}, {Lilly}, {Miniati}, {Bernet}, {Beck},
  {O'Sullivan}, \& {Gaensler}}]{2016ApJ...829..133K}
{Kim}, K.~S., {Lilly}, S.~J., {Miniati}, F., {et~al.} 2016, \apj, 829, 133

\bibitem[{{Kormendy} {et~al.}(2009){Kormendy}, {Fisher}, {Cornell}, \&
  {Bender}}]{2009ApJS..182..216K}
{Kormendy}, J., {Fisher}, D.~B., {Cornell}, M.~E., \& {Bender}, R. 2009, \apjs,
  182, 216

\bibitem[{{Kraljic} {et~al.}(2018){Kraljic}, {Arnouts}, {Pichon}, {Laigle}, {de
  la Torre}, {Vibert}, {Cadiou}, {Dubois}, {Treyer}, {Schimd}, {Codis}, {de
  Lapparent}, {Devriendt}, {Hwang}, {Le Borgne}, {Malavasi}, {Milliard},
  {Musso}, {Pogosyan}, {Alpaslan}, {Bland-Hawthorn}, \&
  {Wright}}]{2018MNRAS.474..547K}
{Kraljic}, K., {Arnouts}, S., {Pichon}, C., {et~al.} 2018, \mnras, 474, 547

\bibitem[{{Lacey} {et~al.}(2016){Lacey}, {Baugh}, {Frenk}, {Benson}, {Bower},
  {Cole}, {Gonzalez-Perez}, {Helly}, {Lagos}, \& {Mitchell}}]{Lacey2016}
{Lacey}, C.~G., {Baugh}, C.~M., {Frenk}, C.~S., {et~al.} 2016, \mnras, 462,
  3854

\bibitem[{{Lacy} {et~al.}(2019){Lacy}, {Baum}, {Chandler}, {Chatterjee},
  {Clarke}, {Deustua}, {English}, {Farnes}, {Gaensler}, {Gugliucci},
  {Hallinan}, {Kent}, {Kimball}, {Law}, {Lazio}, {Marvil}, {Mao}, {Medlin},
  {Mooley}, {Murphy}, {Myers}, {Osten}, {Richards}, {Rosolowsky}, {Rudnick},
  {Schinzel}, {Sivakoff}, {Sjouwerman}, {Taylor}, {White}, {Wrobel}, {Beasley},
  {Berger}, {Bhatnagar}, {Birkinshaw}, {Bower}, {Brand t}, {Brown},
  {Burke-Spolaor}, {Butler}, {Comerford}, {Demorest}, {Fu}, {Giacintucci},
  {Golap}, {Guth}, {Hales}, {Hiriart}, {Hodge}, {Horesh}, {Ivezic}, {Jarvis},
  {Kamble}, {Kassim}, {Liu}, {Loinard}, {Lyons}, {Masters}, {Mezcua},
  {Moellenbrock}, {Mroczkowski}, {Nyland}, {O'Dea}, {O'Sullivan}, {Peters},
  {Radford}, {Rao}, {Robnett}, {Salcido}, {Shen}, {Sobotka}, {Witz}, {Vaccari},
  {van Weeren}, {Vargas}, {Williams}, \& {Yoon}}]{2019arXiv190701981L}
{Lacy}, M., {Baum}, S.~A., {Chandler}, C.~J., {et~al.} 2019, arXiv e-prints,
  arXiv:1907.01981

\bibitem[{{Laing}(1988)}]{Laing88}
{Laing}, R.~A. 1988, \nat, 331, 149

\bibitem[{{Laing} {et~al.}(2008){Laing}, {Bridle}, {Parma}, \&
  {Murgia}}]{Laing2008}
{Laing}, R.~A., {Bridle}, A.~H., {Parma}, P., \& {Murgia}, M. 2008, \mnras,
  391, 521

\bibitem[{{Lan} \& {Prochaska}(2020)}]{LanP2020}
{Lan}, T.-W., \& {Prochaska}, J.~X. 2020, \mnras, 496, 3142

\bibitem[{{Li} {et~al.}(2019){Li}, {Li}, {Bryan}, {Ostriker}, \&
  {Quataert}}]{Miao20192}
{Li}, M., {Li}, Y., {Bryan}, G.~L., {Ostriker}, E.~C., \& {Quataert}, E. 2019,
  arXiv e-prints, arXiv:1909.04204

\bibitem[{{Li} {et~al.}(2020){Li}, {Li}, {Bryan}, {Ostriker}, \&
  {Quataert}}]{Miao2019}
---. 2020, \apj, 894, 44

\bibitem[{{Mac Low} \& {Klessen}(2004)}]{MLK04}
{Mac Low}, M.-M., \& {Klessen}, R.~S. 2004, Reviews of Modern Physics, 76, 125

\bibitem[{{Mac Low} {et~al.}(1998){Mac Low}, {Klessen}, {Burkert}, \&
  {Smith}}]{MacLow98}
{Mac Low}, M.-M., {Klessen}, R.~S., {Burkert}, A., \& {Smith}, M.~D. 1998,
  \prl, 80, 2754

\bibitem[{{Malik} {et~al.}(2020){Malik}, {Chand}, \& {Seshadri}}]{MalikCS2020}
{Malik}, S., {Chand}, H., \& {Seshadri}, T.~R. 2020, \apj, 890, 132

\bibitem[{{Mao} {et~al.}(2017){Mao}, {Carilli}, {Gaensler}, {Wucknitz},
  {Keeton}, {Basu}, {Beck}, {Kronberg}, \& {Zweibel}}]{Mao17}
{Mao}, S.~A., {Carilli}, C., {Gaensler}, B.~M., {et~al.} 2017, Nature
  Astronomy, 1, 621

\bibitem[{{Mathews} \& {Brighenti}(1997)}]{MB97}
{Mathews}, W.~G., \& {Brighenti}, F. 1997, \apj, 488, 595

\bibitem[{{Mathews} \& {Brighenti}(2003)}]{MB03}
---. 2003, Annual Review of Astronomy and Astrophysics, 41, 191

\bibitem[{{Metzger} {et~al.}(2019){Metzger}, {Margalit}, \&
  {Sironi}}]{2019MNRAS.485.4091M}
{Metzger}, B.~D., {Margalit}, B., \& {Sironi}, L. 2019, \mnras, 485, 4091

\bibitem[{{Minter} \& {Spangler}(1996)}]{MS96}
{Minter}, A.~H., \& {Spangler}, S.~R. 1996, \apj, 458, 194

\bibitem[{{Monin} \& {Yaglom}(1971)}]{Monin_Yaglom1971}
{Monin}, A.~S., \& {Yaglom}, A.~M. 1971, {Statistical Fluid Mechanics}
  (Cambridge, Massachusetts, USA: MIT Press)

\bibitem[{{Moss} \& {Shukurov}(1996)}]{MShu96}
{Moss}, D., \& {Shukurov}, A. 1996, \mnras, 279, 229

\bibitem[{{Murgia} {et~al.}(2004){Murgia}, {Govoni}, {Feretti}, {Giovannini},
  {Dallacasa}, {Fanti}, {Taylor}, \& {Dolag}}]{Murgia2004}
{Murgia}, M., {Govoni}, F., {Feretti}, L., {et~al.} 2004, \aap, 424, 429

\bibitem[{{Narasimha} \& {Chitre}(2004)}]{Narasimha2004}
{Narasimha}, D., \& {Chitre}, S.~M. 2004, Journal of Korean Astronomical
  Society, 37, 355

\bibitem[{{Nyland} {et~al.}(2017){Nyland}, {Young}, {Wrobel}, {Davis},
  {Bureau}, {Alatalo}, {Morganti}, {Duc}, {de Zeeuw}, {McDermid}, {Crocker}, \&
  {Oosterloo}}]{Nyland2017}
{Nyland}, K., {Young}, L.~M., {Wrobel}, J.~M., {et~al.} 2017, \mnras, 464, 1029

\bibitem[{{Ogorzalek} {et~al.}(2017){Ogorzalek}, {Zhuravleva}, {Allen},
  {Pinto}, {Werner}, {Mantz}, {Canning}, {Fabian}, {Kaastra}, \& {de
  Plaa}}]{OZ2017}
{Ogorzalek}, A., {Zhuravleva}, I., {Allen}, S.~W., {et~al.} 2017, \mnras, 472,
  1659

\bibitem[{{Ohno} \& {Shibata}(1993)}]{OS93}
{Ohno}, H., \& {Shibata}, S. 1993, \mnras, 262, 953

\bibitem[{{Ostriker} \& {McKee}(1988)}]{OM88}
{Ostriker}, J.~P., \& {McKee}, C.~F. 1988, Reviews of Modern Physics, 60, 1

\bibitem[{{O'Sullivan} {et~al.}(2001){O'Sullivan}, {Forbes}, \&
  {Ponman}}]{OSullivan2001}
{O'Sullivan}, E., {Forbes}, D.~A., \& {Ponman}, T.~J. 2001, \mnras, 328, 461

\bibitem[{{Pacholczyk}(1970)}]{1970ranp.book.....P}
{Pacholczyk}, A.~G. 1970, {Radio astrophysics. Nonthermal processes in galactic
  and extragalactic sources} (San Francisco: Freeman)

\bibitem[{{Patnaik} {et~al.}(1999){Patnaik}, {Kemball}, {Porcas}, \&
  {Garrett}}]{1999MNRAS.307L...1P}
{Patnaik}, A.~R., {Kemball}, A.~J., {Porcas}, R.~W., \& {Garrett}, M.~A. 1999,
  \mnras, 307, L1

\bibitem[{{Penoyre} {et~al.}(2017){Penoyre}, {Moster}, {Sijacki}, \&
  {Genel}}]{2017MNRAS.468.3883P}
{Penoyre}, Z., {Moster}, B.~P., {Sijacki}, D., \& {Genel}, S. 2017, \mnras,
  468, 3883

\bibitem[{{Petroff} {et~al.}(2019){Petroff}, {Hessels}, \&
  {Lorimer}}]{Petroff19}
{Petroff}, E., {Hessels}, J.~W.~T., \& {Lorimer}, D.~R. 2019, \aapr, 27, 4

\bibitem[{{Price} {et~al.}(2011){Price}, {Federrath}, \&
  {Brunt}}]{PriceFederrathBrunt2011}
{Price}, D.~J., {Federrath}, C., \& {Brunt}, C.~M. 2011, \apjl, 727, L21

\bibitem[{{Rincon}(2019)}]{Rincon19}
{Rincon}, F. 2019, Journal of Plasma Physics, 85, 205850401

\bibitem[{{Ruzmaikin} {et~al.}(1988){Ruzmaikin}, {Sokoloff}, \&
  {Shukurov}}]{RSS88}
{Ruzmaikin}, A.~A., {Sokoloff}, D.~D., \& {Shukurov}, A.~M., eds. 1988,
  Astrophysics and Space Science Library, Vol. 133, {Magnetic fields of
  galaxies}

\bibitem[{{Schekochihin} {et~al.}(2002){Schekochihin}, {Cowley}, {Hammett},
  {Maron}, \& {McWilliams}}]{SCHMM02}
{Schekochihin}, A.~A., {Cowley}, S.~C., {Hammett}, G.~W., {Maron}, J.~L., \&
  {McWilliams}, J.~C. 2002, New Journal of Physics, 4, 84

\bibitem[{{Schekochihin} {et~al.}(2004){Schekochihin}, {Cowley}, {Taylor},
  {Maron}, \& {McWilliams}}]{SCTMM04}
{Schekochihin}, A.~A., {Cowley}, S.~C., {Taylor}, S.~F., {Maron}, J.~L., \&
  {McWilliams}, J.~C. 2004, \apj, 612, 276

\bibitem[{{Seta}(2019)}]{Seta2019}
{Seta}, A. 2019, PhD thesis, Newcastle University, Newcastle Upon Tyne, UK.
\newblock \url{http://theses.ncl.ac.uk/jspui/handle/10443/4685}

\bibitem[{{Seta} \& {Beck}(2019)}]{SB2019}
{Seta}, A., \& {Beck}, R. 2019, Galaxies, 7, 45

\bibitem[{{Seta} {et~al.}(2015){Seta}, {Bhat}, \& {Subramanian}}]{SBS2015}
{Seta}, A., {Bhat}, P., \& {Subramanian}, K. 2015, Journal of Plasma Physics,
  81, 395810503

\bibitem[{{Seta} {et~al.}(2020){Seta}, {Bushby}, {Shukurov}, \&
  {Wood}}]{Seta19}
{Seta}, A., {Bushby}, P.~J., {Shukurov}, A., \& {Wood}, T.~S. 2020, Physical
  Review Fluids, 5, 043702

\bibitem[{{Seta} {et~al.}(2018){Seta}, {Shukurov}, {Wood}, {Bushby}, \&
  {Snodin}}]{SSWBS18}
{Seta}, A., {Shukurov}, A., {Wood}, T.~S., {Bushby}, P.~J., \& {Snodin}, A.~P.
  2018, \mnras, 473, 4544

\bibitem[{{Shukurov}(2004)}]{Shukurov2004}
{Shukurov}, A. 2004, ArXiv Astrophysics e-prints, astro-ph/0411739

\bibitem[{{Shukurov} {et~al.}(2017){Shukurov}, {Snodin}, {Seta}, {Bushby}, \&
  {Wood}}]{SSSBW17}
{Shukurov}, A., {Snodin}, A.~P., {Seta}, A., {Bushby}, P.~J., \& {Wood}, T.~S.
  2017, \apjl, 839, L16

\bibitem[{{Shukurov} \& {Sokoloff}(2007)}]{SS2008}
{Shukurov}, A., \& {Sokoloff}, D. 2007, in Les Houches, Session LXXXVIII,
  Dynamos, ed. P.~{Cardin} \& L.~F. {Cugliandolo}, Vol.~88 (Amsterdam:
  Elsevier), 251--299.
\newblock \url{http://dx.doi.org/10.1016/S0924-8099(08)80008-X}

\bibitem[{{Sokoloff} {et~al.}(1998){Sokoloff}, {Bykov}, {Shukurov},
  {Berkhuijsen}, {Beck}, \& {Poezd}}]{Sokoloff1998}
{Sokoloff}, D.~D., {Bykov}, A.~A., {Shukurov}, A., {et~al.} 1998, \mnras, 299,
  189

\bibitem[{{Statler}(2012)}]{St12}
{Statler}, T.~S. 2012, in Astrophysics and Space Science Library, Vol. 378,
  Astrophysics and Space Science Library, ed. D.-W. {Kim} \& S.~{Pellegrini},
  207

\bibitem[{{Stone} {et~al.}(1998){Stone}, {Ostriker}, \& {Gammie}}]{Stone98}
{Stone}, J.~M., {Ostriker}, E.~C., \& {Gammie}, C.~F. 1998, \apjl, 508, L99

\bibitem[{{Strom} \& {Jaegers}(1988)}]{1988A&A...194...79S}
{Strom}, R.~G., \& {Jaegers}, W.~J. 1988, \aap, 194, 79

\bibitem[{{Subramanian}(1999)}]{Sub99}
{Subramanian}, K. 1999, \prl, 83, 2957

\bibitem[{{Subramanian} {et~al.}(2006){Subramanian}, {Shukurov}, \&
  {Haugen}}]{SSH06}
{Subramanian}, K., {Shukurov}, A., \& {Haugen}, N.~E.~L. 2006, \mnras, 366,
  1437

\bibitem[{{Sur}(2019)}]{Sur19}
{Sur}, S. 2019, \mnras, 488, 3439

\bibitem[{{Taylor} {et~al.}(2015){Taylor}, {Agudo}, {Akahori}, {Beck},
  {Gaensler}, {Heald}, {Johnston-Hollitt}, {Langer}, {Rudnick}, {Scaife},
  {Schleicher}, {Stil}, \& {Ryu}}]{Taylor2015}
{Taylor}, R., {Agudo}, I., {Akahori}, T., {et~al.} 2015, in Advancing
  Astrophysics with the Square Kilometre Array (AASKA14), 113

\bibitem[{{Tobias} {et~al.}(2011){Tobias}, {Cattaneo}, \& {Boldyrev}}]{TCB2011}
{Tobias}, S.~M., {Cattaneo}, F., \& {Boldyrev}, S. 2011, ArXiv e-prints,
  arXiv:1103.3138

\bibitem[{{Tzeferacos} {et~al.}(2018){Tzeferacos}, {Rigby}, {Bott}, {Bell},
  {Bingham}, {Casner}, {Cattaneo}, {Churazov}, {Emig}, {Fiuza}, {Forest},
  {Foster}, {Graziani}, {Katz}, {Koenig}, {Li}, {Meinecke}, {Petrasso}, {Park},
  {Remington}, {Ross}, {Ryu}, {Ryutov}, {White}, {Reville}, {Miniati},
  {Schekochihin}, {Lamb}, {Froula}, \& {Gregori}}]{Tz18}
{Tzeferacos}, P., {Rigby}, A., {Bott}, A.~F.~A., {et~al.} 2018, Nature
  Communications, 9, 591

\bibitem[{{van Dokkum} {et~al.}(2010){van Dokkum}, {Whitaker}, {Brammer},
  {Franx}, {Kriek}, {Labb{\'e}}, {Marchesini}, {Quadri}, {Bezanson},
  {Illingworth}, {Muzzin}, {Rudnick}, {Tal}, \& {Wake}}]{2010ApJ...709.1018V}
{van Dokkum}, P.~G., {Whitaker}, K.~E., {Brammer}, G., {et~al.} 2010, \apj,
  709, 1018

\bibitem[{{Wilkin} {et~al.}(2007){Wilkin}, {Barenghi}, \&
  {Shukurov}}]{Wilkin2007}
{Wilkin}, S.~L., {Barenghi}, C.~F., \& {Shukurov}, A. 2007, \prl, 99, 134501

\bibitem[{{York} {et~al.}(2000){York}, {Adelman}, {Anderson}, {Anderson},
  {Annis}, {Bahcall}, {Bakken}, {Barkhouser}, {Bastian}, {Berman}, {Boroski},
  {Bracker}, {Briegel}, {Briggs}, {Brinkmann}, {Brunner}, {Burles}, {Carey},
  {Carr}, {Castander}, {Chen}, {Colestock}, {Connolly}, {Crocker}, {Csabai},
  {Czarapata}, {Davis}, {Doi}, {Dombeck}, {Eisenstein}, {Ellman}, {Elms},
  {Evans}, {Fan}, {Federwitz}, {Fiscelli}, {Friedman}, {Frieman}, {Fukugita},
  {Gillespie}, {Gunn}, {Gurbani}, {de Haas}, {Haldeman}, {Harris}, {Hayes},
  {Heckman}, {Hennessy}, {Hindsley}, {Holm}, {Holmgren}, {Huang}, {Hull},
  {Husby}, {Ichikawa}, {Ichikawa}, {Ivezi{\'c}}, {Kent}, {Kim}, {Kinney},
  {Klaene}, {Kleinman}, {Kleinman}, {Knapp}, {Korienek}, {Kron}, {Kunszt},
  {Lamb}, {Lee}, {Leger}, {Limmongkol}, {Lindenmeyer}, {Long}, {Loomis},
  {Loveday}, {Lucinio}, {Lupton}, {MacKinnon}, {Mannery}, {Mantsch}, {Margon},
  {McGehee}, {McKay}, {Meiksin}, {Merelli}, {Monet}, {Munn}, {Narayanan},
  {Nash}, {Neilsen}, {Neswold}, {Newberg}, {Nichol}, {Nicinski}, {Nonino},
  {Okada}, {Okamura}, {Ostriker}, {Owen}, {Pauls}, {Peoples}, {Peterson},
  {Petravick}, {Pier}, {Pope}, {Pordes}, {Prosapio}, {Rechenmacher}, {Quinn},
  {Richards}, {Richmond}, {Rivetta}, {Rockosi}, {Ruthmansdorfer}, {Sandford},
  {Schlegel}, {Schneider}, {Sekiguchi}, {Sergey}, {Shimasaku}, {Siegmund},
  {Smee}, {Smith}, {Snedden}, {Stone}, {Stoughton}, {Strauss}, {Stubbs},
  {SubbaRao}, {Szalay}, {Szapudi}, {Szokoly}, {Thakar}, {Tremonti}, {Tucker},
  {Uomoto}, {Vanden Berk}, {Vogeley}, {Waddell}, {Wang}, {Watanabe},
  {Weinberg}, {Yanny}, {Yasuda}, \& {SDSS Collaboration}}]{2000AJ....120.1579Y}
{York}, D.~G., {Adelman}, J., {Anderson}, Jr., J.~E., {et~al.} 2000, \aj, 120,
  1579

\bibitem[{{Zeldovich} {et~al.}(1990){Zeldovich}, {Ruzmaikin}, \&
  {Sokoloff}}]{ZRS90}
{Zeldovich}, {\relax Ya}.~B., {Ruzmaikin}, A.~A., \& {Sokoloff}, D.~D. 1990,
  {The Almighty Chance} (Singapore: World Scientific)

\bibitem[{{Zweibel}(2020)}]{Zweibel2020}
{Zweibel}, E.~G. 2020, \apj, 890, 67

\end{thebibliography}

\end{document}